\documentclass[usenatbib]{mn2e}
\usepackage{amsmath,graphicx}
\begin{document}
\title[Formation of Star Clusters]{The Formation of Star Clusters II: 3D Simulations of Magnetohydrodynamic Turbulence in Molecular Clouds}
\author[Tilley, D. A. and Pudritz, R. E.]{David A. Tilley$^1$\thanks{E-mail: dtilley@nd.edu (DAT); pudritz@physics.mcmaster.ca (REP)} and Ralph E. Pudritz$^{2,3}$\footnotemark[1] \\ $^1$ Department of Physics, University of Notre Dame, Notre Dame, Indiana, USA 46556 \\ $^2$ Department of Physics and Astronomy, McMaster University, Hamilton, Ontario, Canada L8S 4M1 \\ $^3$ Origins Institute, McMaster University, Hamilton, Ontario, Canada L8S 4M1}
\maketitle
\begin{abstract}
  We present a series of decaying turbulence simulations that
  represent a cluster-forming clump within a molecular cloud,
  investigating the role of magnetic fields on the formation of
  potential star-forming cores.  We present an exhaustive analysis of
  numerical data from these simulations that includes a compilation of
  all of the distributions of physical properties that characterize
  bound cores - including their masses, radii, mean densities, angular
  momenta, spins, magnetizations, and mass-to-flux ratios.  We also
  present line maps of our models that can be compared with
  observations.  Our simulations range between 5-30 Jeans masses of
  gas, and are representative of molecular cloud clumps with masses
  between $100-1000 M_\odot$.  The field strengths in the bound cores
  that form tend to have the same ratio of gas pressure to magnetic
  pressure, $\beta$, as the mean $\beta$ of the simulation.  The cores
  have mass-to-flux ratios that are generally less than that of the
  original cloud, and so a cloud that is initially highly
  supercritical can produce cores that are slightly supercritical,
  similar to that seen by Zeeman measurements of molecular cloud
  cores.  Clouds that are initially only slightly supercritical will
  instead collapse along the field lines into sheets, and the cores
  that form as these sheets fragment have a different distribution of
  masses than what is observed.  The spin rates of these cores
  (wherein 20-40\% of cores have $\Omega t_{ff}\ge 0.2$) suggests that
  subsequent fragmentation into multiple systems is likely.  The sizes
  of the bound cores that are produced are typically 0.02-0.2 pc and
  have densities in the range $10^4-10^5\;\mathrm{cm}^{-3}$ in
  agreement with observational surveys.  Finally, our numerical data
  allow us to test theoretical models of the mass spectrum of cores,
  such as the turbulent fragmentation picture of Padoan-Nordlund.  We
  find that while this model gets the shape of the core mass spectrum
  reasonably well, it fails to predict the peak mass in the core mass
  spectrum.
\end{abstract}

\section{Introduction}

Molecular clouds are highly dynamic environments and understanding the
effects that their internal, supersonic turbulent motions have on the
process of gravitational collapse is critical towards understanding
how stars form.  The turbulent fragmentation picture of star formation
posits that star formation is regulated by the turbulent motions of
the molecular cloud; turbulent fluctuations can support the cloud on
large scale while also creating compressions that can become
gravitationally unstable, wherein protostellar cores form after 1-2
free-fall times; 0.5-1 Myr (eg., reviews by \citep{klessen_maclow04,
  elmegreen_etal00}).  This is in stark contrast with the traditional
paradigm of star formation, which supposes that a prestellar core in a
magnetically subcritical cloud contracts quasistatically as the
magnetic field slowly leaks out due to ambipolar diffusion on
time-scales of several Myr (e.g.  \cite*{mestel_spitzer56,
  mouschovias_paleologou81, paleologou_mouschovias83, shu_etal87,
  ciolek_basu01}).

Our goal in this and the previous paper (\citeauthor{tilley_pudritz04}
\citeyear{tilley_pudritz04}, TP04 hereafter) is to understand the
origin of star clusters, in particular the origin of the distribution
of molecular cloud cores (and their physical properties) out of which
cluster stars are formed.  Numerical investigations of turbulence
within GMCs have shown that turbulent fragmentation is remarkably
successful in producing clusters of stars \citep{gammie_ostriker96,
  klessen_etal00, heitsch_etal01, ostriker_stone_gammie01,
  padoan_nordlund02, li_etal04, klessen_etal05, vs_etal05}.  Many of
the observed statistical properties of cluster-forming regions in
molecular clouds can be reproduced by these models, including:

\noindent
(i) {\it the apparent relation between internal velocity dispersions
  and size} \citep{burkert_bodenheimer00, ostriker_stone_gammie01,
  bp_maclow02, ossenkopf_maclow02, heyer_brunt04};

\noindent
(ii) {\it the similarity with the observed distribution of core
  masses}
(\cite{klessen_burkert_bate98,klessen_burkert00,balsara_etal01b,heitsch_etal01,klessen01a,klessen01b,klessen_burkert01,reipurth_clarke01,bate_bonnell_bromm02,bp_maclow02,bate_bonnell_bromm03,li_etal03,nordlund_padoan03,gammie_etal03,ostriker03,clark_bonnell04,dd_clarke_bate04,dd_etal04b,klessen04,li_etal04,padoan_nordlund04}; TP04; \cite{bate_bonnell05,jappsen_etal05,martel_evans_shapiro06,padoan_etal05,bp_etal06})
- wherein the numerical data shares many of the key features of
observed star-forming region, such as the (Salpeter) $\sim -1.35$
power-law relationship at large masses between number of cores of a
given mass and the core mass and a turnover at low masses
\citep{kramer_etal98,motte_andre_neri98,testi_sargent98,jijina_etal99,johnstone_etal00,luhman00,luhman_etal00,johnstone_etal01,briceno_etal02,luhman_etal03,reid_wilson05};

\noindent
(iii)
{\it the similarity of the core mass function to the stellar initial mass function (IMF)} 
(\citep{miller_scalo79,scalo86,scalo98,kroupa02,chabrier03});  

\noindent
(iv) {\it the similarity of the angular momentum of cores with
  observed rotations of cores} (\cite{burkert_bodenheimer00,
  klessen_burkert00, gammie_etal03, fisher04, jappsen_klessen04,
  li_etal04},TP04)- which are consistent with observations of velocity
gradients in molecular clouds \citep{goodman_etal93,
  barranco_goodman98, caselli_etal02b, caselli_etal02a}, and several
orders of magnitude greater than the angular momenta of main-sequence
stars; and

\noindent
(v) {\it the moderate efficiency of star formation within dense
  clumps} wherein a fraction of the gas in these models collects in
the cores, up to 30-40\% in models of decaying turbulence
(\cite{gammie_etal03, clark_bonnell04, vs_etal03};TP04 ;
\cite{clark_etal05}).  Models with driven turbulence predict lower star formation efficiencies (e.g. \cite{vs_etal03}), and it has been suggested that the star formation needs to be spread out over several dynamical times \citep{krumholz_mckee05, tan_krumholz_mckee06, krumholz_tan07}.

Our previous work on purely hydrodynamic turbulence in
self-gravitating gas (TP04) found that only a few initial Jeans masses
are needed to create a cluster of prestellar cores.  This is because
the shocks and compressions create local density enhancements with
much reduced local Jeans masses (see also \citet{goodwin_etal04a,
  goodwin_etal04b, goodwin_etal04c, clark_bonnell05}).  The point here
is that since the local Jeans mass is inversely proportional to the
square root of the density, it can become significantly easier to form
collapsing cores in these density enhancements.

In the present paper, we extend our investigation to include the role
that magnetic fields can play in turbulent fragmentation.  Magnetic
fields can play a role in principle because of the well known fact
that they change the condition for gravitational stability.  In a
uniform magnetized medium, the condition for collapse is most
physically measured by the magnetic criticality parameter (or
mass-to-flux ratio) $\Gamma = 2\pi\sqrt{G}\Sigma/B$, where $\Sigma$ is
a surface density of the gas and B is its magnetic field strength.  In
situations in which the magnetic field is weak compared to gravity --
the so-called magnetically supercritical case, $\Gamma > 1$ --
gravitational collapse is expected.  The subcritical configuration
($\Gamma < 1$) has magnetic fields strong enough to support the cloud
against gravitational collapse, and is thus expected to be stable.  A
condensation can collapse only if the magnetic field can escape via
ambipolar diffusion, reducing the magnetic flux and thus increasing
$\Gamma$ (e.g. \cite*{mestel_spitzer56,mouschovias_paleologou81}).
\citet{li_nakamura04} have demonstrated that it is possible for
turbulence to enhance this diffusion rate, thus hastening the
transition from subcritical to supercritical behaviour, but it is not
clear if the cores formed from this type of process resemble GMC
cores.

Given that the unmagnetized simulations have had such success - what
role does the magnetic field play?  Based on the previous paragraph,
it is obvious that in regions within the cloud wherein the field is
too strong (and couples ideally to the gas), magnetic forces can
prevent collapse from occurring.  This has been confirmed in
simulations of subcritical MHD turbulence (e.g.
\cite*{heitsch_etal01,ostriker_stone_gammie01}), although it has been
suggested that the magnetic field can leak out of condensations via
turbulence-enhanced ambipolar diffusion in reasonable time-scales,
producing supercritical cores out of subcritical regions
\citep{fatuzzo_adams02,kim_diamond02,zweibel02,li_nakamura04}.  If the
initial cloud is magnetically supercritical, collapse will occur.
Another important consideration that is not well handled by idealized
analytical models of clouds is the fact that geometry of the magnetic
field is important since the field does not resist motions parallel to
it.  A third factor that makes magnetic fields potentially interesting
is that any rotation within the cores can generate torsional Alfv\'en
waves that can carry away excess angular momentum and possibly
generate jets while affecting the fragmentation of the disk that will
ultimately form \citep{mouschovias_paleologou79,
  mouschovias_paleologou80, blandford_payne82, pudritz_norman83,
  lovelace_berk_contopoulos91, basu_mouschovias94}.  Do these
important effects influence the efficiency of forming bound cores in
turbulence?

In this paper, we examine the effect of a magnetic field in addition
to turbulence in self-gravitating, cluster-forming clumps within GMCs.
We perform an exhaustive examination of the contributions to the
virial equation from the thermal, kinetic, gravitational, magnetic and
surface terms of the fluctuations that are produced by the turbulence.
We use the virial equation to identify the objects that are bound or
collapsing, which we identify as 'cores'.  The cores are confined
through a combination of thermal, dynamic and magnetic surface
pressure, and gravity.  The virial equation provides a useful guide to
determine the forces acting on the cores.  In a turbulent environment,
the surface pressure terms can have a significant role in confining
the core (\cite{mckee_zweibel92,bp_vs_scalo99},TP04).  An apparent
equipartition, at least in a statistical sense, develops between the
internal kinetic energy and the gravitational energy
(TP04,\cite{klessen_etal05}).

We find that strong initial magnetic fields ($\Gamma \approx 1-5$) can indeed
have a large effect on the resulting evolution of the cloud and 
result in collapse to large sheets.  There is little evidence for such
prevalent structures from observations however, and we use this to
constrain the physical parameters of our simulations to uncover the
conditions that are most favourable to the formation of observed star
clusters.  The simulations that initially have large $\Gamma$ appear
to do much better at reproducing the observed trends seen in molecular
cloud mass functions.  These simulations can still produce cores with
local values of  
$\Gamma$ that are close to critical, as the cores are produced when the
turbulence breaks up the fluid into parcels with smaller masses and
magnetic fluxes than the original computational domain.  We also find
good agreement between our core properties with ammonia surveys of
cores in clustered star formation regions (e.g.
\cite{jijina_etal99}).

We also use our numerical data to compute the angular momenta and spin
rates for the cores in our simulations.  We find that the angular
momenta are on the order of $10^{22-23}\;\mathrm{cm^2 s^{-1}}$,
comparable to measurements of GMC cores
\citep{goodman_etal93,barranco_goodman98,caselli_etal02b,caselli_etal02a}.
The distribution of core spins that we find suggests to us that the
binary frequency of stars that form within these objects, must be
high.

This work, wherein we compare our results with 
a wide variety of observations of cloud cores, shows that the strong magnetization
of cores that is often observed is the result of local compression by the turbulence, and is not
characteristic of the entire volume of the cloud.  This and other results suggest that it is
turbulence, and not the wide-spread influence of a powerful 
cloud magnetic field, that is central to the origin of core formation and core properites and ultimately, the origin of the
IMF.

The setup of our simulations is described in Section 2, 
and focuses mainly on the differences from the simulations in TP04.  
In Section 3, we describe the overall results of the simulations, 
noting especially the morphological structure that is produced.  
In Section 4, we examine the dynamical state of the cores and 
condensations that form.  We present, in Section 5, the distribution functions 
of many physical properties of cores, including their masses, angular momenta,
spins, radii, average densities, magnetizations and mass-to-flux ratios. We compare
these with the observations and use them to test theoretical models of turbulent
fragmentation.


\section{Simulations}\label{section_simulations}

\begin{table*}\begin{center}
\begin{tabular}{r|rrrrrrrr}
\hline Run & Spectrum & $n_j$ & $m_\mathrm{tot}/m_\odot$ & L/pc & $M\dagger$ & $\beta$ & $M_A\ddagger$ & $\Gamma$\\
B5b & K & 4.6 & 105.1 & 0.32 & 5.0 & 0.9 & 4.7 &4.9\\
B5c & K & 4.6 & 105.1 & 0.32 & 5.0 & 5.0 & 11.2 &11.6\\\vspace{1mm}
B5d & K & 4.6 & 105.1 & 0.32 & 5.0 & 10.0 & 15.8 &16.4\\
C5c & K & 7.5 & 453.9 & 1.0 & 5.0 & 5.0 & 11.2 &13.7\\
C5d & K & 7.5 & 453.9 & 1.0 & 5.0 & 10.0 & 15.8 &19.3\\\vspace{1mm}
C5e & K & 7.5 & 453.9 & 1.0 & 5.0 & 50.0 & 35.4 &43.2\\
D5a & K & 12.0 & 623.7 & 1.0 & 5.0 & 0.1 & 1.6 &2.3\\
D5b & K & 12.0 & 623.7 & 1.0 & 5.0 & 1.0 & 5.0 &7.1\\\vspace{1mm}
D5c & K & 12.0 & 623.7 & 1.0 & 5.0 & 3.9 & 9.9 &14.1\\
E14b& B & 27.5 & 1086.3 & 1.0 & 14.1 & 1.0 & 14.1 &9.4\\
\hline\multicolumn{6}{l}{$\dagger$ $M=v_\mathrm{RMS}/c_s$}\\
\multicolumn{6}{l}{$\ddagger$ $M_A = v_\mathrm{RMS}/v_A$}
\end{tabular}\end{center}
\caption{Initial conditions for the simulations presented in this paper. The first letter specifies the number of Jeans masses -- 'B' for $n_J=4.6$, 'C' for 7.5, 'D' for 12.0, and 'E' for 27.5.  The number represents the initial RMS thermal Mach number of the simulation, '$\mathcal{M}$'.  The final letter denotes the mean $\beta$ of the simulation, where $\beta$ is the ratio of thermal pressure to magnetic pressure.  'a' represents $\beta=0.1$, 'b' represents $\beta\approx 1$, 'c' represents $\beta\approx4-5$, 'd' represents $\beta=10$, and 'e' represents $\beta=50$.\label{table_ic}}
\end{table*}

\begin{figure}
\begin{center}
\includegraphics[width=84mm]{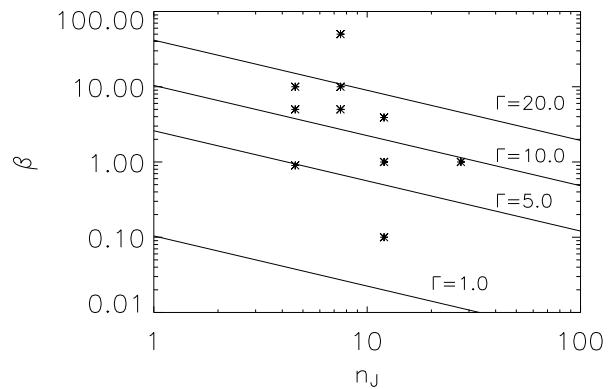}
\caption{Initial conditions for the simulations presented in this paper.  The various data points correspond to the values in Table \ref{table_ic}.  Superimposed are lines of constant magnetic critical number $\Gamma$.\label{fig_simic}}
\end{center}
\end{figure}
\subsection{Initial Conditions}

The simulations presented here are set up in a very similar manner to those in TP04, and are described in more detail there.  We use the \textsc{zeus-mp} code \citep{stone_norman92a,stone_norman92b,norman00}, a magnetohydrodynamic fluid dynamics code made available by NCSA.  MHD forces are calculated using the Method of Characteristics \citep{hawley_stone95}; self-gravity of the fluid is calculated using the \textsc{FFTW} fast fourier transform libraries \citep{frigo_johnson98}.  Our boundary conditions are set to a periodic grid, and represent the centre of a GMC.  The calculations presented here were performed on a Compaq AlphaServer SC40 at the SHARCNET McMaster University site.  These simulations have a resolution of $256^3$ grid cells, and were typically simulated using 8 processors.  As the cores we wish to simulate have densities of $10^4-10^9\;\mathrm{cm}^{-3}$, an isothermal equation of state is appropriate \citep{hayashi66,larson69}.  We have gravity active from the beginning of our simulations, but this does not have a significant effect until the turbulence has significantly decayed.

We start our simulations with an initially uniform density field.  We initiate the turbulence by creating a set of plane waves in Fourier space, with relative amplitudes drawn from either a $\gamma=-5/3$ 1D Kolmogorov spectrum (marked 'K' in Table \ref{table_ic}) or a $\gamma=-2$ 1D Burgers spectrum (marked 'B' in Table \ref{table_ic}) such that the velocity fluctuations as a function of 3D wavenumber $k=2\pi/\lambda$ scale as $v_k^2 \propto k^{\gamma-2}$.  We remove waves longer than 1/8 the length of the box in order to minimize the effects of the periodicity of the box on the kinematics.  We also imposed an exponential cutoff on short wavelengths, to eliminate fluctuations in the initial conditions on scales we are not able to properly resolve. This spectrum is illustrated in a later figure.  The initial velocity field is purely solenoidal, but otherwise each plane wave is given a random phase and direction.  We do not provide any forcing of the turbulence at later timesteps; the initial kinetic energy is allowed to decay freely.  After $\sim 1$ shock-crossing time, the turbulence is fully developed.

The initial magnetic field is uniform in strength and direction.  We characterize its initial strength through the ratio of thermal energy to magnetic energy, 
\begin{eqnarray}\beta=4\pi\rho c_s^2/B^2\end{eqnarray}
We select a range of $\beta$ between 0.1 and 50.0 (the Zeeman measurements of \citet{crutcher99} suggest $\beta \approx 0.1$); the initial parameter space of $(n_J,\beta)$ is plotted in Fig. (\ref{fig_simic}).

The list of our various model parameters is given in Table \ref{table_ic}, where models with initial Jeans number $n_J = 4.6, 7.5, 12.0$ and $27.5$ are designated with letters 'B','C','D' or 'E' respectively.  The following number is the RMS Mach number of the initial turbulence spectrum; all of our simulations save one (E14b) were run with a turbulent Mach number $\mathcal{M}=v_\mathrm{RMS}/c_s=5$.  We chose this value because experience with our hydrodynamic studies (which featured models similar to our 'B' models) showed this to be a good regime for vigorous turbulent fragmentation into a well-populated CMF.  The final, lower-case letter gives an indication of the strength of the initial magnetic field, with $\beta\sim 0.1,1,5,10,50$ indicated respectively by a,b,c,d,e.

The turbulent amplitude can also be viewed in terms of the Alfv\'en Mach number, $\mathcal{M}_A = v_\mathrm{RMS}/v_A$ where $v_A = B/\sqrt(4\pi\rho) = c_s/\sqrt{\beta}$ is the signal speed of MHD transverse waves.  The values of this number are given by 
\begin{eqnarray}\mathcal{M}_A = \sqrt{\beta} \mathcal{M}\end{eqnarray}
which for $\mathcal{M}=5$ and our range of $\beta$ gives $1.6\le M_A\le35$.  Thus, all the simulations in this paper were performed in the mildly to strongly super-Alfv\'enic regime.  The initial level of turbulent kinetic energy is greater than the gravitational self-energy of the cloud, but as the turbulence decays the gravitational force will begin to dominate the dynamics.

Another important characteristic of the magnetic field that is useful in determining the behaviour of the fluid is whether it is magnetically supercritical or subcritical.  As already noted, a magnetically supercritical fluid has sufficient mass for gravity to overwhelm magnetic support and go into collapse; a magnetically subcritical fluid does not have enough mass, and thus will not collapse gravitationally.  The critical mass to flux ratio -- which is the ratio of the gravitational to magnetic energies in a parcel of gas of density $\rho$, size $L$ and magnetic field strength $B$ -- can be rewritten as
\begin{eqnarray}
\Gamma=\frac{2\pi G^{1/2} \rho L}{B}=3.12\beta^{1/2}n_J^{1/3}\label{eq_gamma_def}
\end{eqnarray}
 \citep{basu_ciolek04}.  We plot our initial simulation models in our 2-D parameter space ($\beta$,$n_J$) in Fig. (\ref{fig_simic}), where the solid lines correspond to constant values of $\Gamma$. The value $\Gamma=1.0$ is the dividing line between supercritical and subcritical magnetic fields and collapse will occur for $\Gamma > 1$.
  The scaling of the critical parameter in Equation (\ref{eq_gamma_def}) with our two fundamental parameters, $\beta$ and $n_J$, shows that strongly magnetized cores ($\beta\ll 1$) can still be supercritical if the initial number of Jeans masses $n_J$ is sufficiently large ($n_J > 0.36 \beta^{-3/2}$).  Thus, sufficient gravity (i.e. large $n_J$) can still leave a strongly magnetized cloud supercritical.

Our simulations were run until we could not properly resolve the collapse everywhere on the grid, a process that generally took about 3 flow-crossing times to occur.  When scaled to the initial values of L in Table \ref{table_ic}, this is $\sim$0.5 Myr, consistent with observations that star formation generally lasts less than a few Myr \citep{hartmann01,hartmann_etal01}.  We used the local Jeans length as our criterion for this resolution, calculated for each cell using the density of that cell.  The simulation stopped when the local Jeans length of a cell somewhere on the grid was less than four zones \citep{truelove_etal97} (this is equivalent to an increase in density of $4148.6n_J^{-2/3}$ over the initial density of the simulation; for our simulation values of $n_J\in[4.6,27.5]$ this corresponds to final-to-initial density ratios of 450-1500).  This criterion was established for simulations in the absence of magnetic fields; \citet{heitsch_etal01} have suggested a more stringent criterion of resolving the Jeans length by at least six zones when magnetic fields are included.  However, our results should not be contaminated by artificial fragmentation due to our less stringent criterion as we stop the simulations immediately upon the violation of this criterion, and the collapse of the region that violates this condition is happening on much faster time-scales than the dynamics anywhere else on the computational grid.  As a result, any instabilities that might result due to violation of the more stringent Heitsch et al. criterion will not have a chance to propagate to other cells before the simulation ends.

\subsection{Comparison of Units with Observations}
  These simulations are scale-free, and can thus, in theory at least, be scaled to any size or mass.  However, it is not necessarily physical to do so.  As in TP04 we chose the initial conditions from among the range of clumps and cores seen in the \citet{lada_etal91} survey; these are cores with masses between $10-500 M_\odot$ and sizes between $0.05-0.5$ pc; they contain $0.5-30$ Jeans masses, assuming they are at a temperature of 20 K.  This is roughly the range we expect our simulations to scale while still being representative of clumps and cores; at larger scales, we would expect clouds to have many more Jeans masses.

The mass, size and magnetic field of our simulations scale as
\begin{eqnarray}
\frac{m_\mathrm{tot}}{m_\odot} & = & 119 n_j^{2/3} \left(\frac{L}{\mathrm{pc}}\right) \left(\frac{T}{20\;\mathrm{K}}\right)\\
\frac{B}{\mu G} & = & 16.8 n_J \beta^{-1/2} \left(\frac{m_\mathrm{tot}}{100 m_\odot}\right)^{-1}
\end{eqnarray}

\subsection{Identification of Cores}

We identify individual condensations in our simulations through the application of a watershed algorithm \citep{vincent_soille91,mangan_whitaker99} that we developed in TP04.  This algorithm efficiently identifies cores by identifying the local gradient vector at each point.  A path is then traced, from zone to zone, following the local gradient vector of each cell until a local maximum is reached.  All zones along this path are assigned to a core marked by that local maximum.  See TP04 for details, including a comparison with the \textsc{clumpfind} algorithm of \citet{williams_etal94}.  

As the algorithm by default is sensitive to small fluctuations, it tends to break up larger cores into multiple smaller cores.  We reduce the influence of this effect on our results by applying the core-finding algorithm to the density field, smoothed with a Gaussian kernel with a width of 3 zones.  This significantly reduces the number of small-scale cores that we find.


\section{Cloud Structure}\label{section_cloud}

\subsection{Sheets, Filaments and Cores}

The structures that form in our simulations are shown in Figs. \ref{magfield_b5c}-\ref{magfield_e14a}, and are delineated by isodensity contours and magnetic field lines.  We find that there is a clear trend in the results for the fluid to flow preferentially along the magnetic field lines, rather than perpendicular to the fields \citep{padoan_nordlund99,li_etal04}.  This effect is naturally more significant in the simulation runs with stronger magnetic fields (i.e. lower $\beta$ and, in particular, low $\Gamma$).  
The simulations with a mean magnetic field strength that is closer to the critical magnetic field strength show that fluid motions perpendicular to the mean magnetic field are reduced, while motions parallel to the mean field are not significantly affected.
In this regime the material forms a sheet that then fragments.  The formation of a sheet is a well-known behaviour of subcritical cores \citep{mouschovias_spitzer76,heitsch_etal01}, although such clouds would not be able to collapse in the absence of ambipolar diffusion \citep{balsara_etal01a,li_nakamura04,vs_etal05}.  However, we have a sheet forming in a marginally supercritical cloud.  We still show in Sections (\ref{section_formation}) and (\ref{section_distribution}) that the cores that form within this sheet have different statistical properties from observed cores and are unlikely to be representative of star-forming regions.

The fluid in the low $\beta$ simulations initially clumps together locally, forming many small condensations.  These condensations are gravitationally attracted to each other, but the component of the motion perpendicular to the field lines is not generally strong enough to exert a significant perpendicular velocity unless the magnetic field is strongly subcritical.  These fluid condensations will instead flow along the field lines, until the gravitational force parallel to the field is reduced as a sheet forms.  As this sheet builds up in mass, the gravitational forces transverse to the field become strong enough to cause local collapse in several parts of the sheet, that then experience the typical runaway growth of a gravitationally collapsing object. This results in several massive cores at the point where we are forced to stop the simulation due to violation of the Jeans criterion, as opposed to the non-magnetic and strongly subcritical runs where one object tends to runaway before any other condensations are significantly evolved.

Furthermore, the separation of the simulation into a high-density sheet and low-density envelope divides the condensations found into two distinct categories: those with a high average density (in the sheet), and those with a low average density (in the atmosphere).   We show in Section (\ref{section_distribution}) that these two sets have distinct properties; as a result, these sheets do not look like places where clusters of stars will form.

\subsection{Magnetic Field Structure in Turbulent Clouds}
\begin{figure*}
\includegraphics[width=158mm]{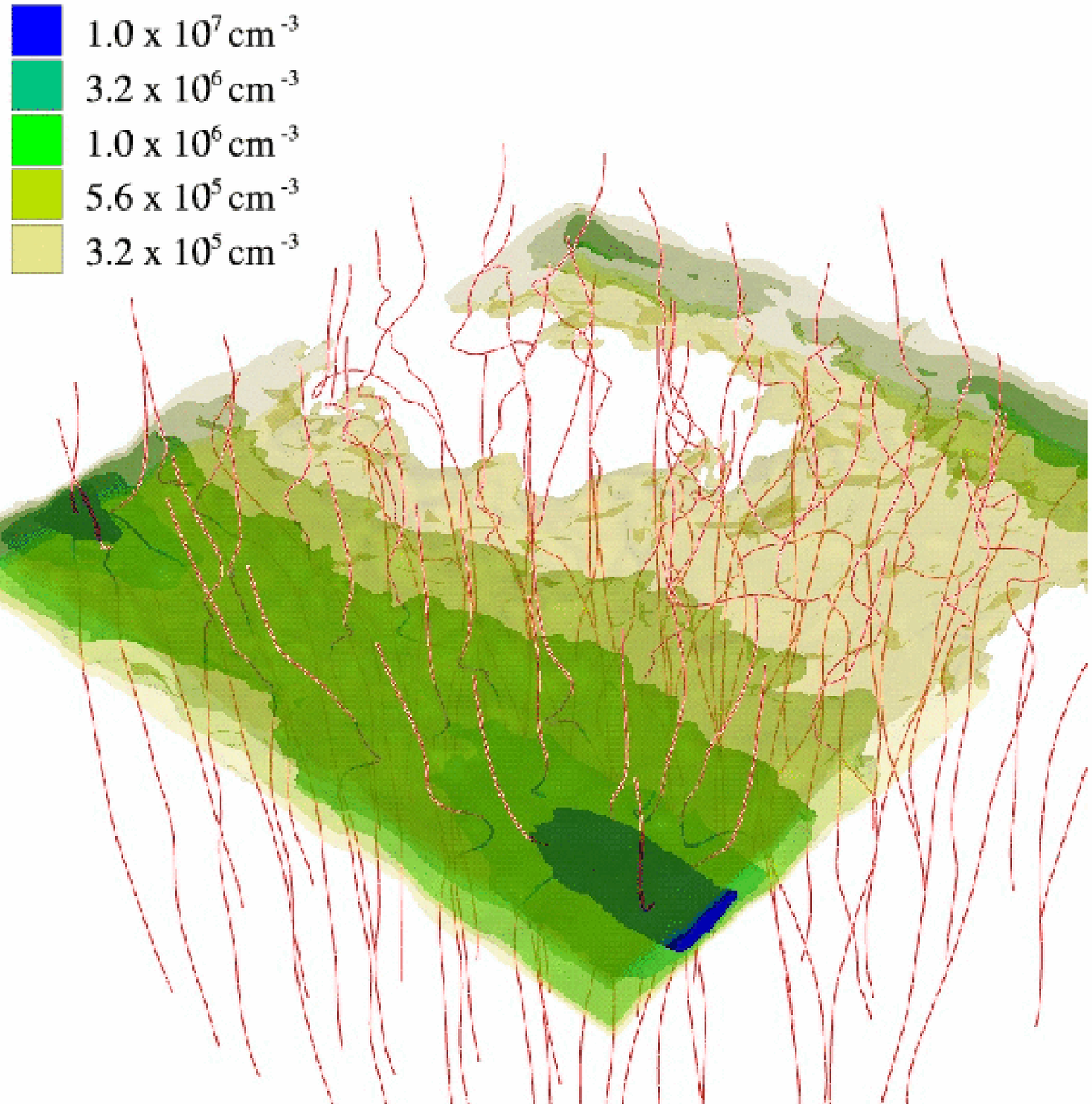}
\caption{Isodensity surfaces and magnetic field configuration for the entire volume of Run B5c.  The edge of the grid has a length of 0.32 pc when scaled to our default values in Table \ref{table_ic}.\label{magfield_b5c}}
\end{figure*}

\begin{figure*}
\includegraphics[width=158mm]{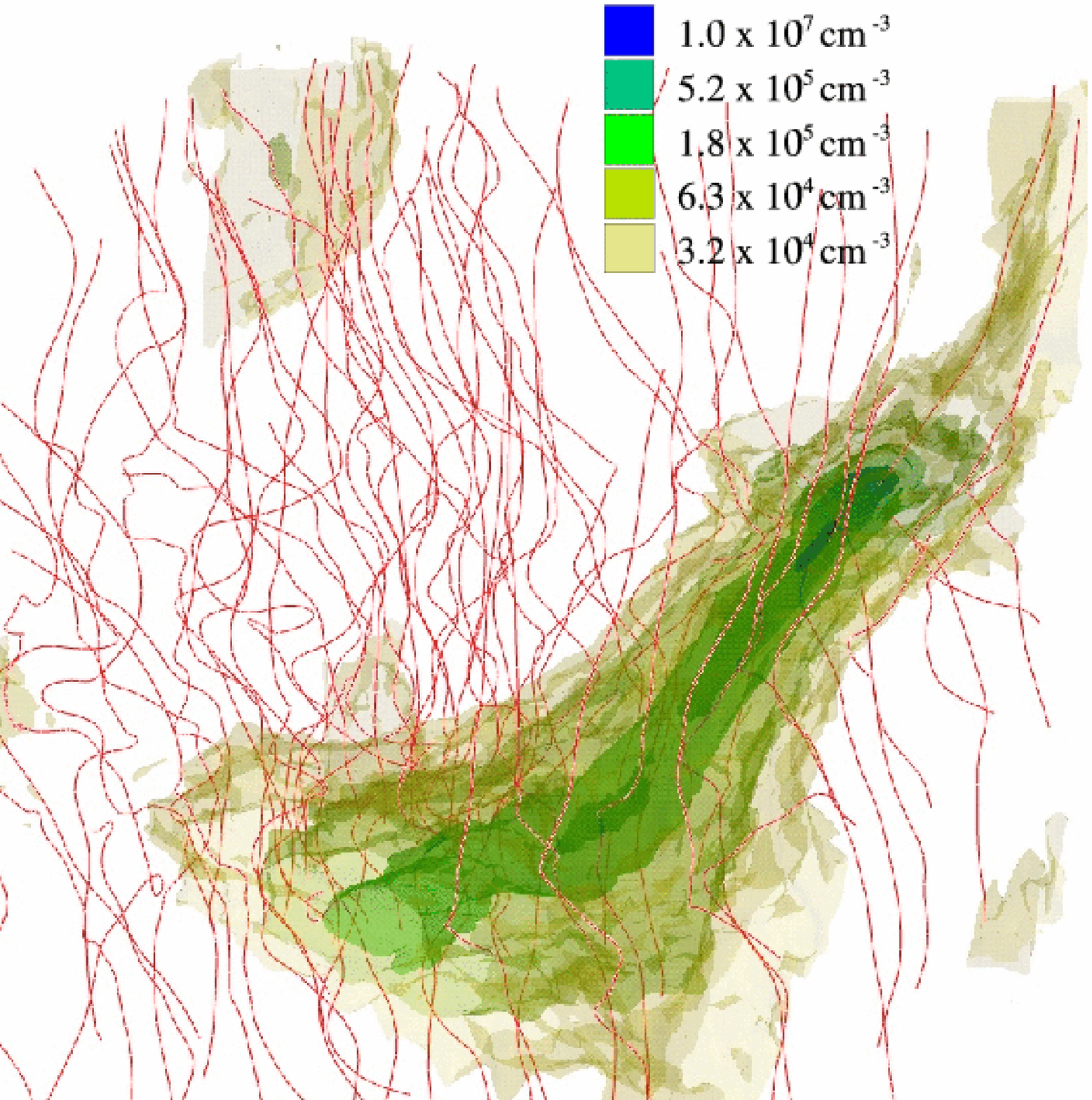}
\caption{Isodensity surfaces and magnetic field configuration for the entire volume of Run C5d.  The edge of the grid has a length of 1.0 pc when scaled to our default values in Table \ref{table_ic}.\label{magfield_c5d}}
\end{figure*}
\begin{figure*}
\includegraphics[width=158mm]{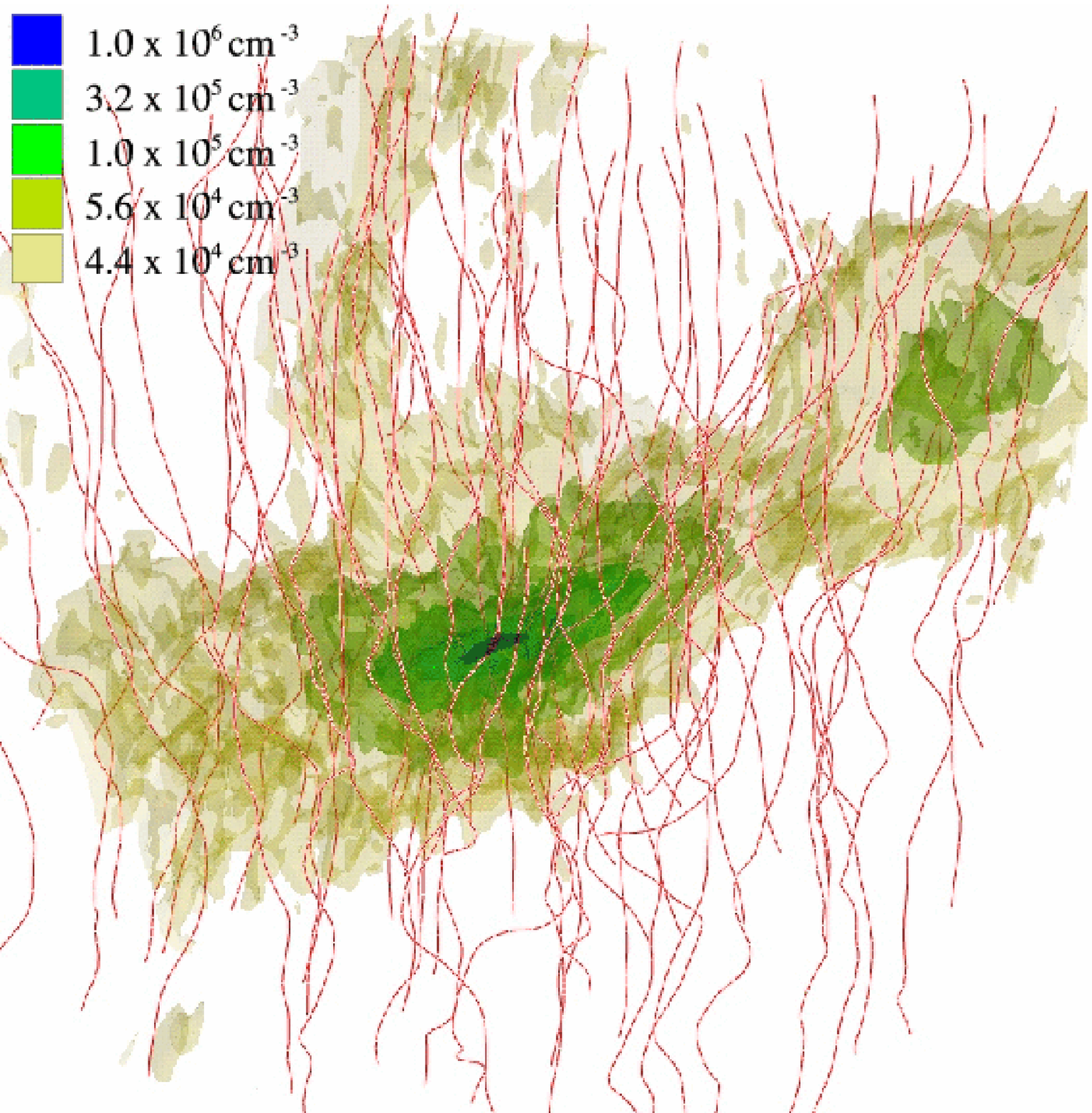}
\caption{Isodensity surfaces and magnetic field configuration for the entire volume of Run D5c.  The edge of the grid has a length of 1.0 pc when scaled to our default values in Table \ref{table_ic}.\label{magfield_d5c}}
\end{figure*}
\begin{figure*}
\includegraphics[width=158mm]{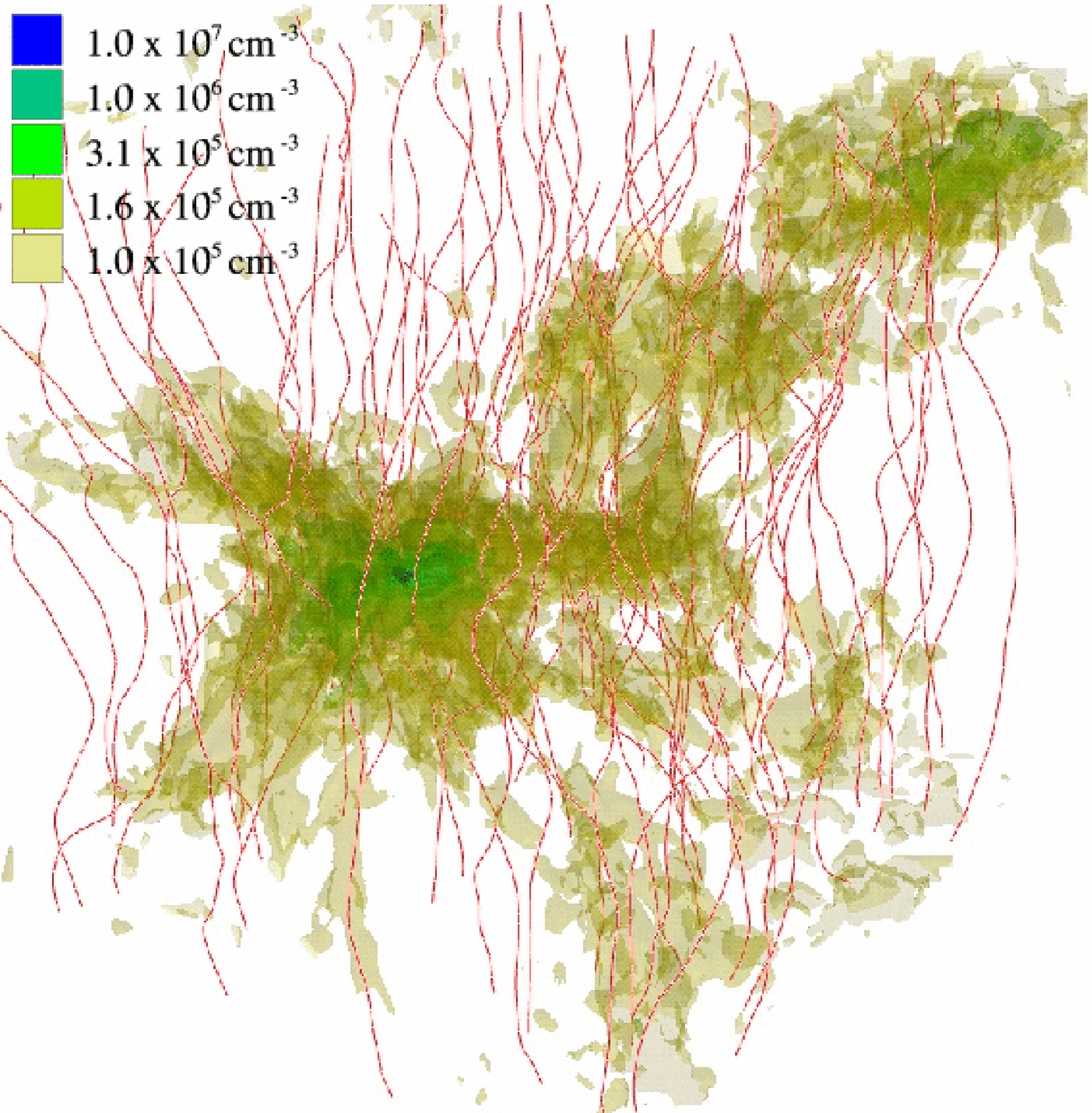}
\caption{Isodensity surfaces and magnetic field configuration for the entire volume of Run E14b.  The edge of the grid has a length of 1.0 pc when scaled to our default values in Table \ref{table_ic}.\label{magfield_e14a}}
\end{figure*}
\begin{figure*}
\includegraphics[width=158mm]{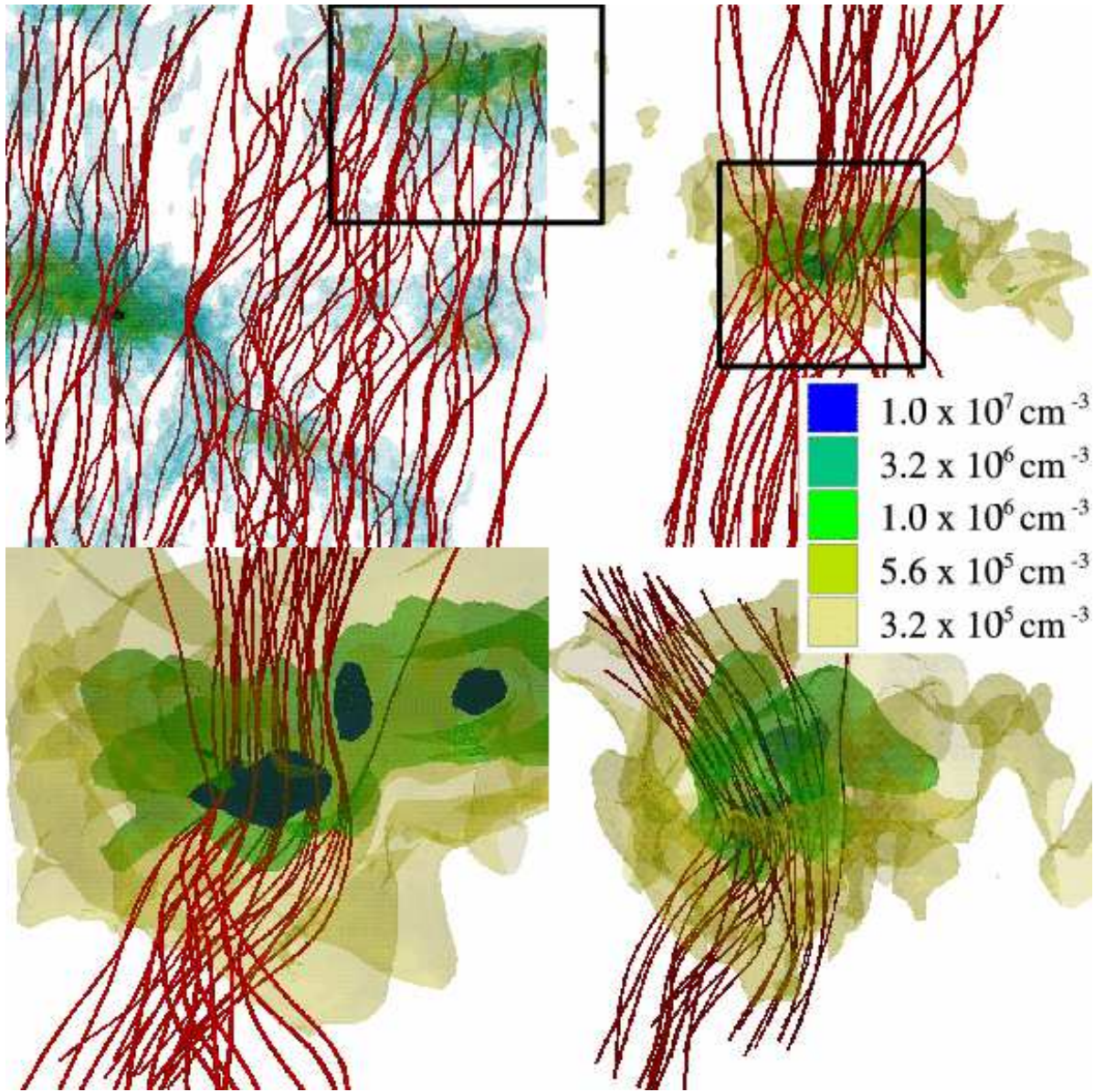}
\caption{Zooming in on the third most massive core in Run E14b.  The top-left panel shows the entire simulation, the top-right panel zooms in by a factor of 2, the lower-left panel zooms in by a further factor of 8/3, and the lower-right panel has the same magnification as the lower-left, but rotated by about 90 degrees to show the field line structure.\label{magfield_closeup}}
\end{figure*}

\begin{figure*}
\includegraphics[width=84mm]{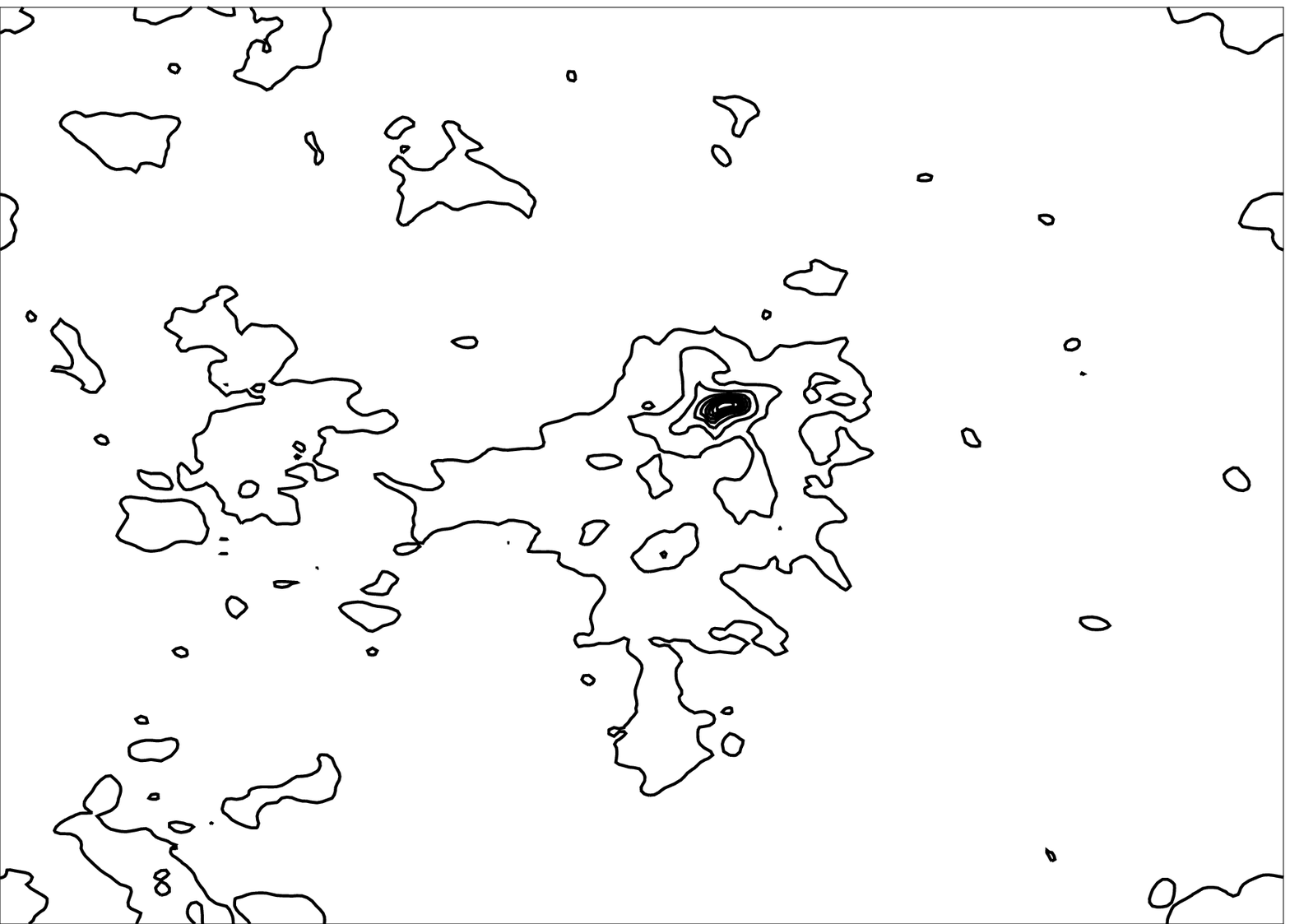}\includegraphics[width=84mm]{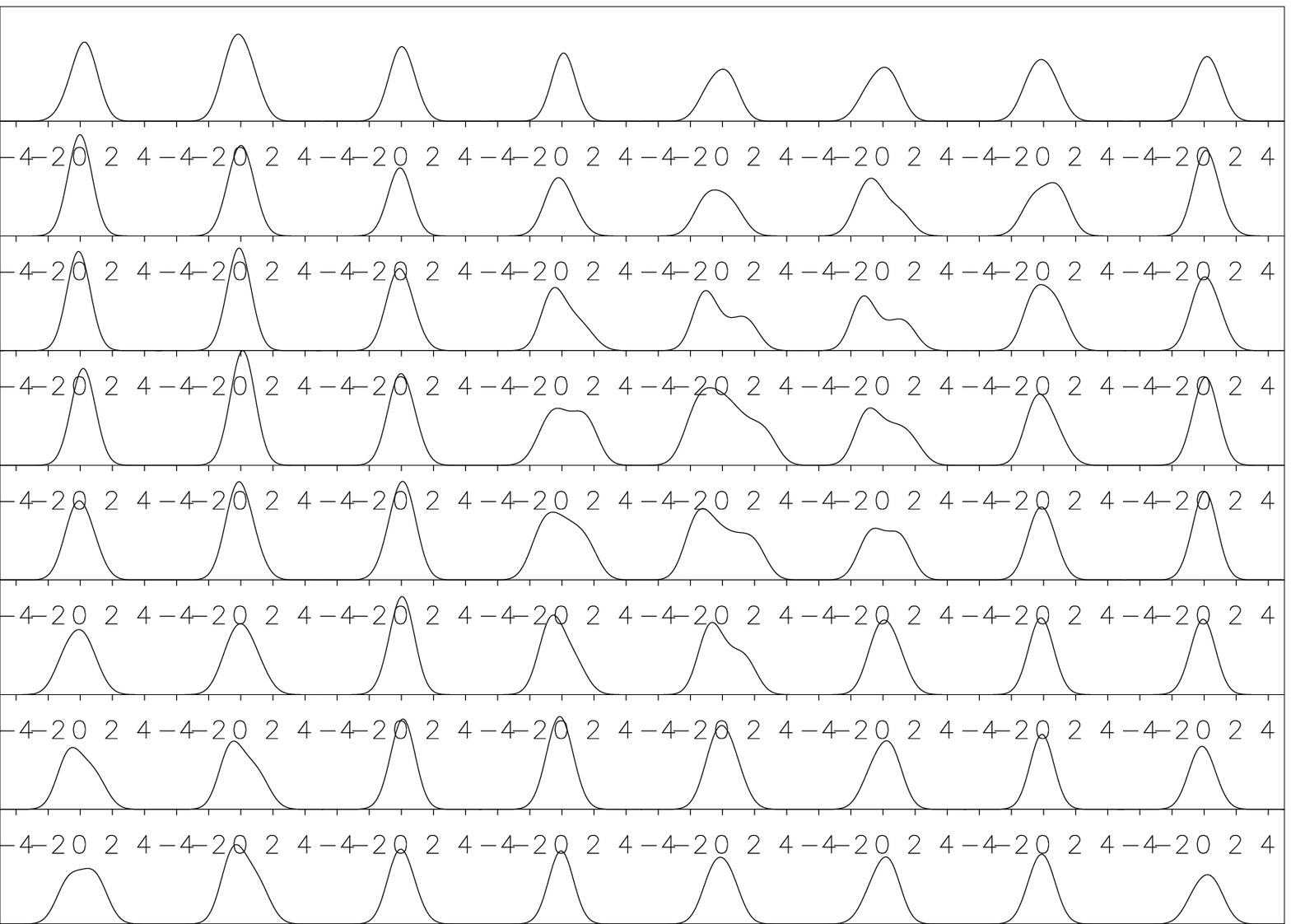}
\includegraphics[width=84mm]{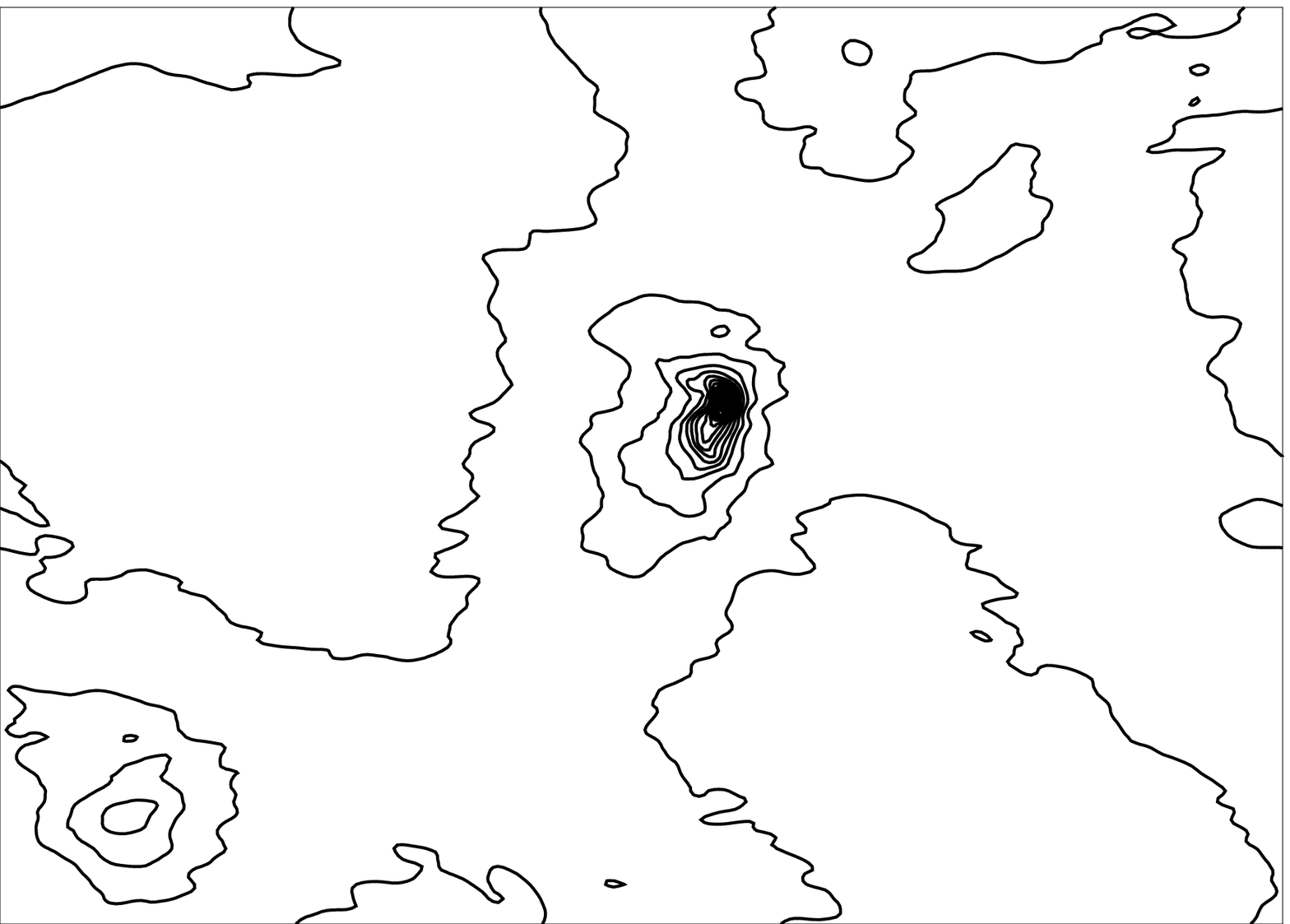}\includegraphics[width=84mm]{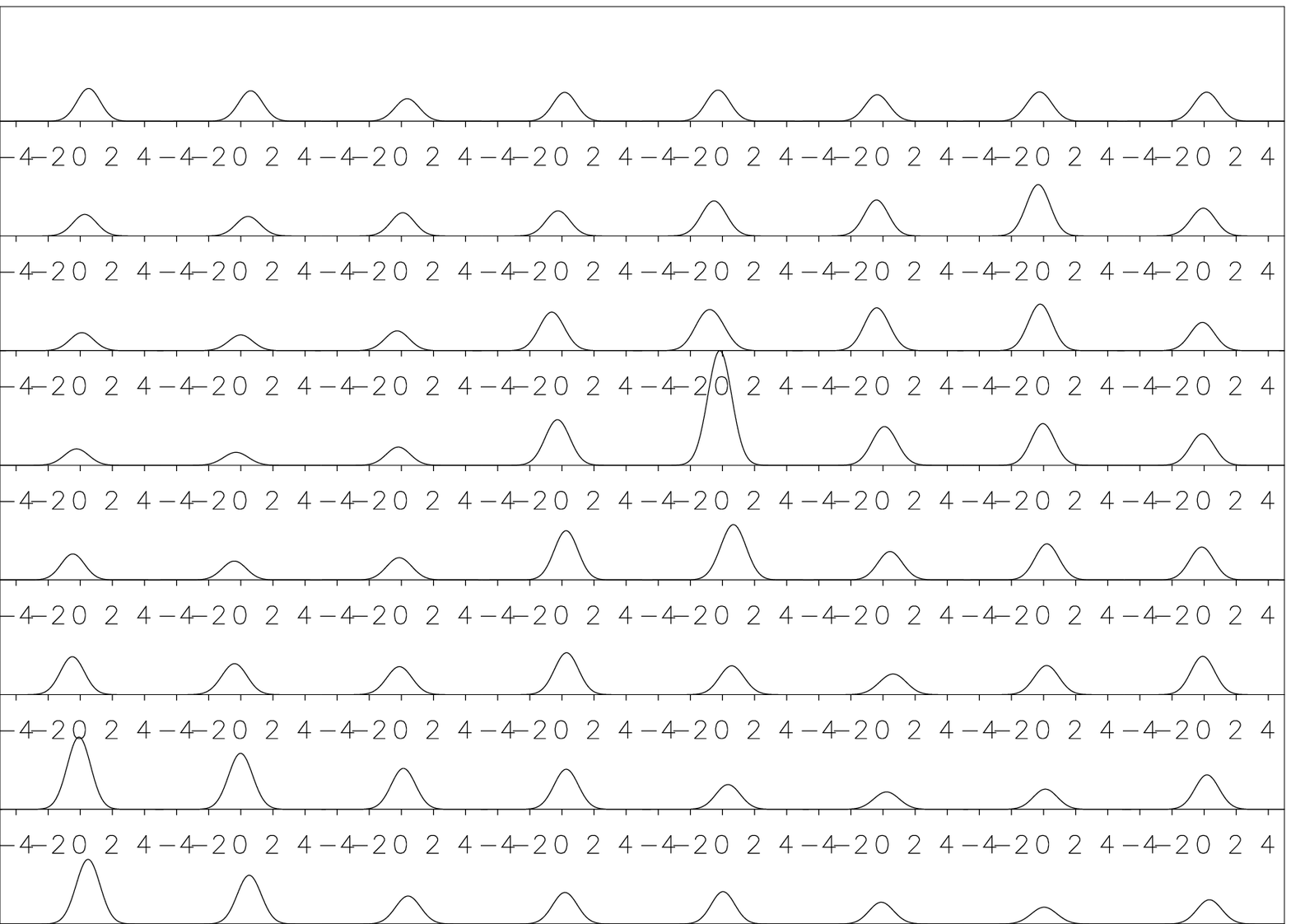}
\caption{Simulated column density contours and line maps of Run E14b, the simulation presented in this paper that most resembles a star-forming cluster.  The top row contains a projection along the mean field axis (left panel) and the accompanying line map (right panel).   The bottom row contains a projection perpendicular to the mean field axis, with the mean field in the horizontal direction. See Fig. \ref{magfield_e14a} for a 3D projection of this simulation.\label{fig_linemap}  The abscissas of the individual line map plots are in units of the thermal sound speed.}
\end{figure*}

The magnetic field in a turbulent molecular cloud can become quite
complex, as shown for two of our runs, C5d and E14b, in Figs.
\ref{magfield_c5d} and \ref{magfield_e14a}.  Oblique shocks generate
vortical motions in the gas which in turn drive torsional Alfv\'en
waves that propagate through the fluid, resulting in the twisting and
tangled structures we see in these two figures.  These propagating
torsional waves create regions of helical magnetic field as described
in theoretical work by \citet{fiege_pudritz00} and which may have been
observed in Zeeman observations \citep{robishaw_heiles05}.  The field
lines can get quite twisted and tangled, even with a relatively strong
magnetic field, as shown in Figs.
\ref{magfield_b5c}-\ref{magfield_closeup}.  The net direction of the
initial field is readily apparent, as magnetic flux is conserved to
numerical accuracy.

We also show a series of close-up views around the third most massive core (zooming in by factors of 2 and 8/3) in Run E14b in Fig. \ref{magfield_closeup}, which is typical of the mid-mass isolated bound cores produced in these simulations.  We chose this particular core to show the details around cores that have not begun to experience a runaway collapse.  We can also see the proximity of this core to two smaller, nearby cores.  The field lines around the target core are not strongly tangled.  As there is less turbulent energy on small scales (by virtue of the Kolmogorov spectrum), it is perhaps not too surprising that the fluid motions are unable to tangle up the field to a significant degree in these smaller cores.  No pinching of the field due to gravitational collapse is seen, as the mass-to-flux ratio for this core is $\Gamma=0.9$.  Examining the virial terms we can see that this core is bound primarily due to the presence of a surface pressure; gravity is only responsible for $~25$\% of the virial confinement.  This core contains $\sim 40000$ zones, so it is well-resolved; it has a density ratio between centre and surface of $\sim 9$, with a central density (when scaled to the initial simulation values in Table \ref{table_ic}) corresponding to $8\times10^5\;\mathrm{cm^{-3}}$.  We note that this is not the core that underwent runaway collapse that stopped the simulation, and we do see evidence of focusing in the magnetic field lines there, but it is remarkable that these cores have grown so large without significant motions transverse to the initial field.

Any rotation of cores like these would be expected to generate torsional waves that extract their angular momentum.  We see some evidence for this in the twisting of the field lines beneath the core even though this core has not had enough time to fully rotate.  Moreover, in this particular core, we find that the angular momentum vector $\mathrm{\mathbf{L}}$ and mean magnetic field $\mathrm{\mathbf{B}}$ are misaligned by 45 degrees.  We examine this in greater detail in a subsequent paper, but note that similar studies have suggested that cores do not strongly align with either local or global magnetic field directions \citep{gammie_etal03}.

\subsection{Line Profiles}

We present simulated line profiles along with a column density contour map for Run E14b in Fig. \ref{fig_linemap}.  The top row uses a line-of-sight parallel to the mean field.  The bottom row uses a line-of-sight perpendicular to the mean field, with the mean field in the horizontal direction.  These line profiles (shown in the right-hand panels of Fig. \ref{fig_linemap}) are calculated in the same way as TP04, by summing along each line-of-sight the density value of each zone multiplied by a Gaussian exponential term centered on the line-of-sight velocity of that zone; the width of the Gaussian is determined by the amount of thermal broadening expected for a gas at a temperature of 20K.  These profiles represent optically thin emission lines that emit over the entire range of densities that we have in our simulations, typically $10^4-10^7$ cm$^{-3}$ (ammonia or CS, for instance).  The amplitudes of the lines are scaled relative to the highest-amplitude line.  The abscissa of the line plots are in units of the sound speed.

The range of densities in Run E14b varies over four orders of magnitude, but the range in column density in the projection along the mean field is only a factor of 14 due to some preference for collapse along the field lines.  As a result, the line profiles all tend to have similar amplitudes.  The profiles become noticeably more asymmetric near the regions of highest column density as material streams on to the large cores through channeling by the magnetic field.  This is not seen in the projection perpendicular to the mean field, which has a much wider variation in column density ($\sim 60$) and less motion transverse to the magnetic field.  As a result, the line profiles have a strong Gaussian character, as observed (e.g. \citet{falgarone_phillips90,schneider_etal96,goodman_etal98,park_lee_myers04}).


\section{Formation and Evolution of Cores}\label{section_formation}

\begin{table}
\begin{tabular}{lrrrr}
\hline Run & $n_J$ & $\beta$ & $\Gamma$ & SFE \\
\hline 
B5b& 4.6 & 0.9 & 4.9 & 0.54\\
B5c& 4.6 & 5.0 & 11.6 & 0.76\\
B5d& 4.6 & 10.0 & 16.4 & 0.76\\
C5c& 7.5 & 5.0 & 13.7 & 0.64\\
C5d& 7.5 & 10.0 & 19.3 & 0.64\\
C5e& 7.5 & 50.0 & 43.2 & 0.46\\
D5a& 12.0 & 0.1 & 2.3 & 0.52\\
D5b& 12.0 & 1.0 & 7.1 & 0.56\\
D5c& 12.0 & 3.9 & 14.1 & 0.47\\
E14b& 27.5 & 1.0 & 9.4 & 0.14\\
\hline
\end{tabular}
\caption{The star formation efficiencies for each of the simulations at the time the Jeans condition is violated.  This measures the fraction of the total mass that is contained within gravitationally bound cores.\label{table_sfe}}
\end{table}

\begin{figure*}
\includegraphics[width=168mm]{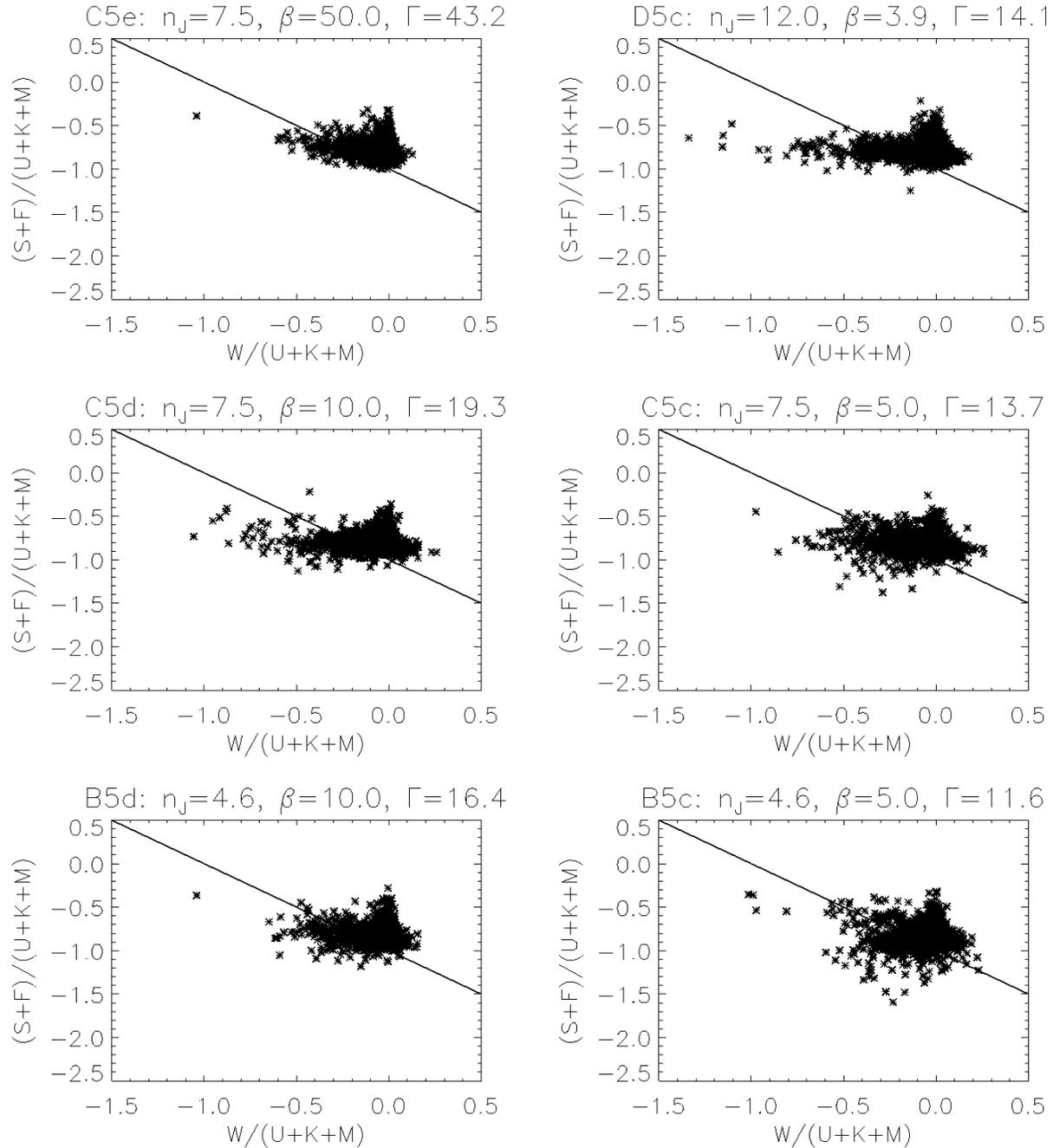}
\caption{The relationship between the various terms in the virial equation in each of the runs shown in Table \ref{table_ic}.  The models are arranged in order of decreasing $\Gamma$.  The surface terms are plotted against the gravitational term, each normalized to the sum of the internal thermal, kinetic and magnetic energy.  The solid line in each plot is $\ddot{I}'=0$.\label{fig_virial}}
\end{figure*}

\begin{figure*}
\includegraphics[width=168mm]{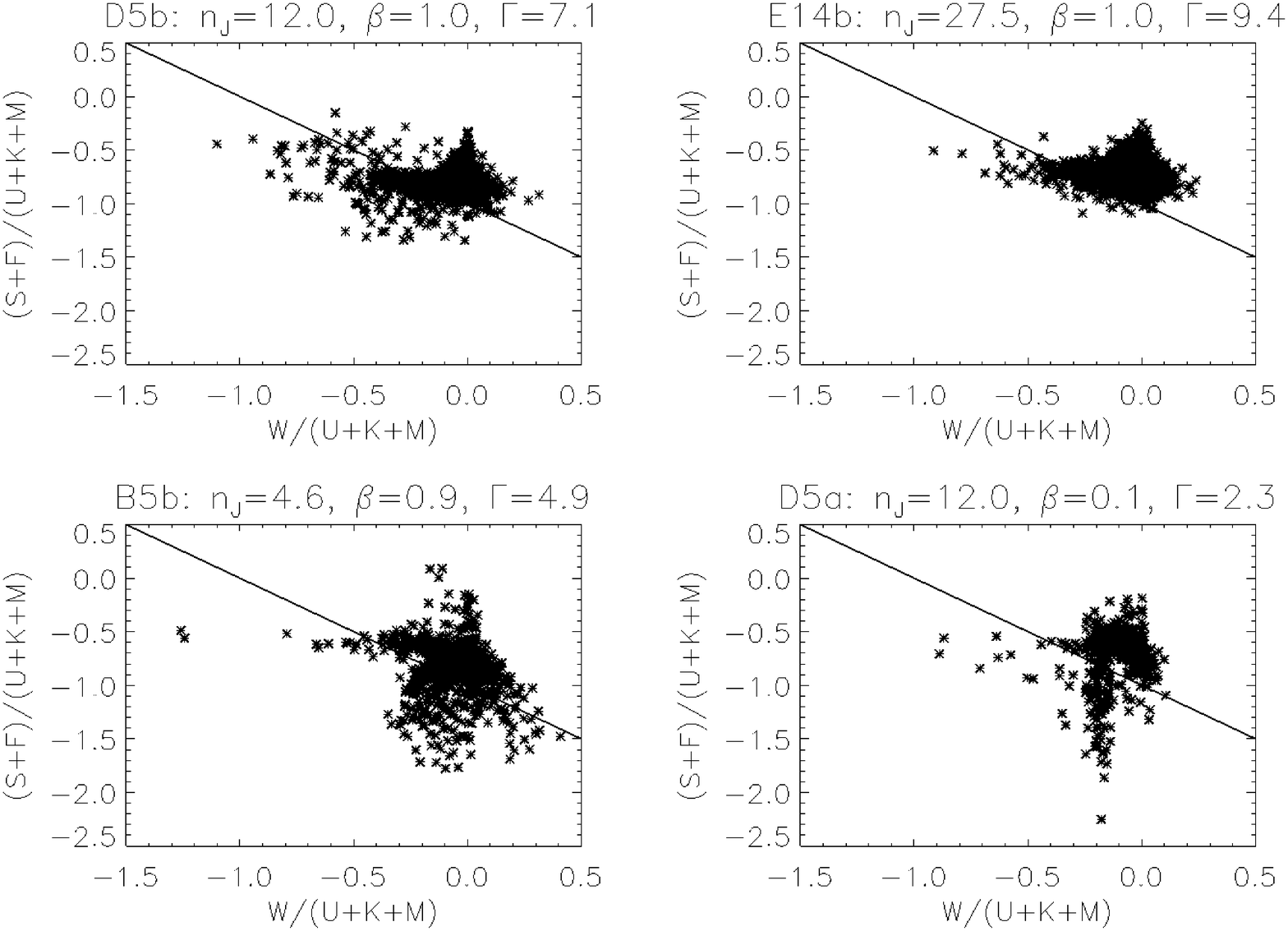}
\contcaption{}
\end{figure*}

\begin{figure*}
\includegraphics[width=84mm]{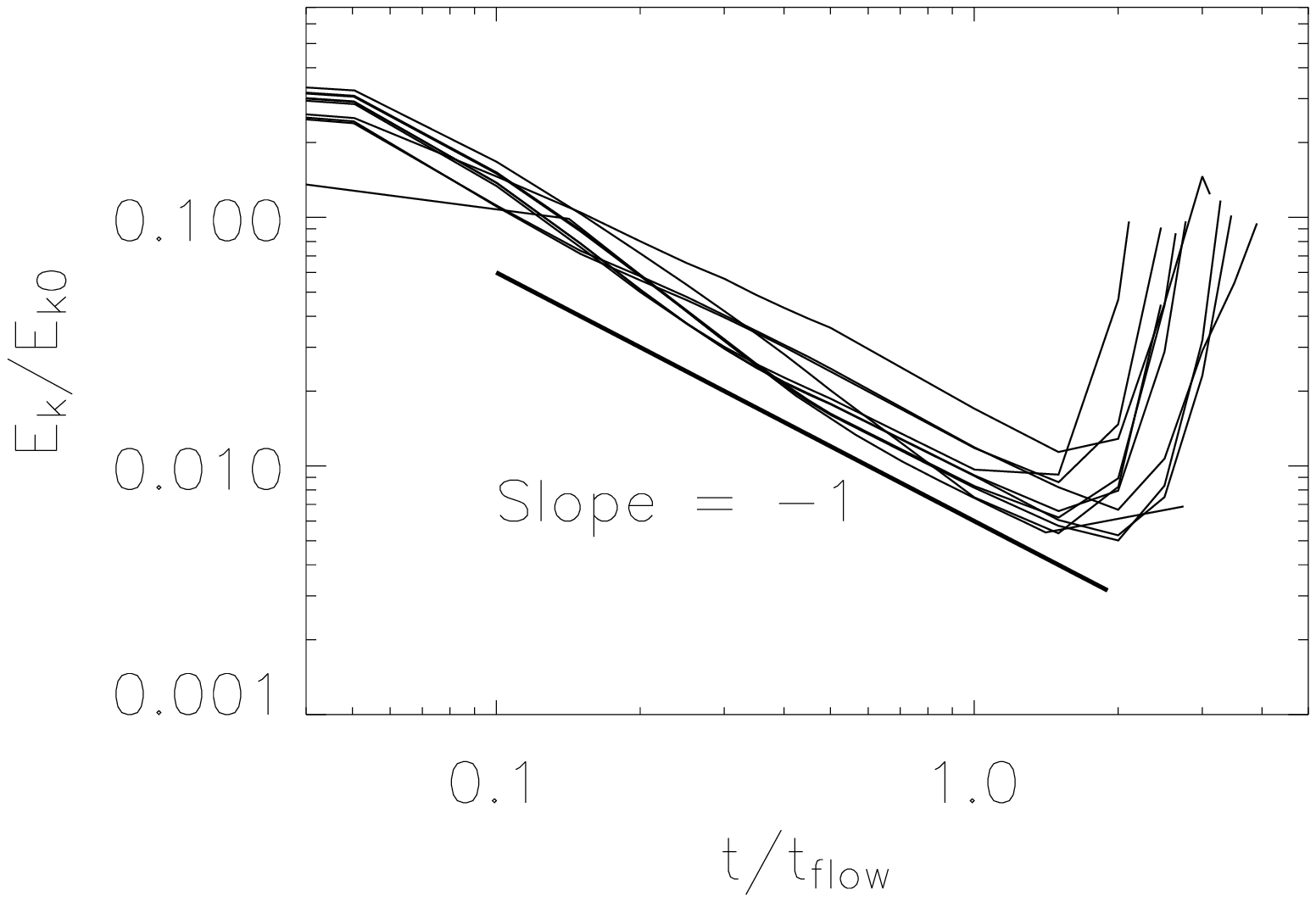}\includegraphics[width=84mm]{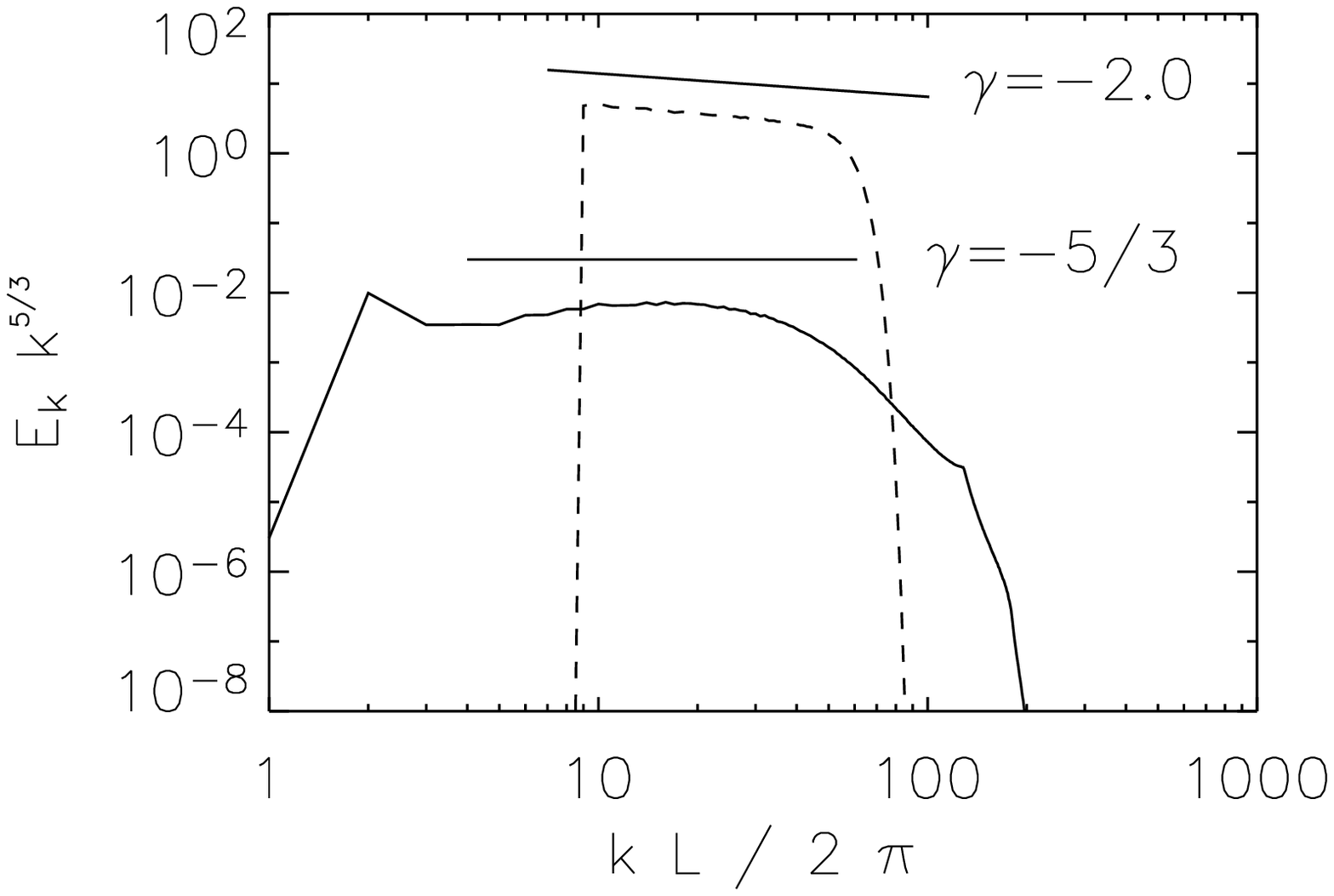}
\caption{(Left) Kinetic energy as a function of time for the different runs presented in this paper.  (Right) The energy spectrum of Run E14b, multiplied by $k^{-5/3}$.  The dashed curve represents the initial spectrum of the turbulence and the solid line the resulting spectrum at the time the Truelove conditions was violated, 7 flow-crossing times later.\label{fig_decay}}
\end{figure*}

\subsection{Energy Balance}

We found in TP04 that an excellent understanding of the dynamical state of the cores was provided by the individual terms in the virial equation.  We apply the virial equation using Eulerian coordinates to the case of a magnetized, self-gravitating fluid as \citep{mckee_zweibel92,bp_vs_scalo99}:

\begin{eqnarray}
\frac{1}{2}\ddot{I} + \frac{1}{2}\int_S \rho r^2 \mathbf{v}\cdot \mathrm{d}\mathbf{S} = U + K + W + S + M + F = \frac{1}{2}\ddot{I}'\label{eq_virial}
\end{eqnarray}
where
\begin{eqnarray}
I & = & \int_V \rho r^2 \mathrm{d}V \\
U & = & 3\int_V P \mathrm{d}V\\
K & = & \int_V \rho v^2 \mathrm{d}V\\
W & = & -\int_V \rho \mathbf{r}\cdot\nabla\Phi\mathrm{d}V\\
S & = & -\int_S\left[P\mathbf{r} + \mathbf{r}\cdot\left(\rho\mathbf{vv}\right)\right]\cdot\mathrm{d}\mathbf{S}\\
M & = & \frac{1}{4\pi}\int_VB^2\mathrm{d}V\\
F & = & \int_S\mathbf{r}\cdot\mathbf{T}_M\cdot\mathrm{d}\mathbf{S}
\end{eqnarray}
where  $\mathbf{v}$ is the velocity, $\rho$ is the density, $P=\rho c_s^2$ is the isothermal pressure, $\Phi$ the gravitational potential, $\mathrm{B}$ the magnetic field, and $\mathbf{T}$ the Maxwell stress-energy tensor.   The terms $I,U,K,W,S,M,F$ indicate respectively the moment of inertia of the fluctuation, the internal thermal energy, the internal kinetic energy, the gravitational energy, the sum of the thermal surface pressure and dynamic surface pressure, the internal magnetic energy, and the magnetic surface pressure.

The volume integrals are calculated over all cells identified as part of the fluctuation.  The surface integrals are calculated from the fluxes through every zone identified as being on the boundary between one fluctuation and the next.

Of these terms, the internal thermal, kinetic and magnetic energies will always be positive; the other terms can be either positive or negative.  We plot the sum of the surface terms S and F against the gravitational term W, each normalized to the sum of the internal terms, in Fig. \ref{fig_virial}.  As in TP04, we label the fluctuations for which $\ddot{I}' < 0$ ``bound cores'', and fluctuations for which $\ddot{I}' > 0$ ``unbound fluctuations''.

To interpret Fig. \ref{fig_virial}, we recall that the virial equation describes a steady state when $\ddot{I}=0$ in Eq. \ref{eq_virial}.  This corresponds to the straight line in Fig. \ref{fig_virial}.  A core that falls on this line is in virial balance, with the support provided by the internal kinetic energy K, thermal energy U and magnetic energy M balancing the confinement provided by the gravitational energy W, the magnetic pressure and tension on the surface F, and the surface thermal+ram pressure S.  A core will collapse if it lies below the line in Fig. \ref{fig_virial}.

The highly supercritical simulations such as Run C5e show a virial plot that is very similar to the plots we found in TP04, where there were typically one or a few cores that were strongly gravitationally dominant, and the bulk of the cores either lying close to the line $\ddot{I}' = 0$ or with $W/(K+U) \sim 0$ (in TP04, there was no magnetic field, and thus M=0). Most of the rest of the bound cores are only marginally virial unstable, while the unbound cores can be quite far from virial stability due to significant internal pressure overwhelming the confining forces.

We see in Fig. \ref{fig_virial} that the surface terms have a more significant contribution to the virial equation as the mean magnetic field becomes stronger.  In particular, the two runs with the smallest $\Gamma$, Run B5b and Run D5a, have surface terms that act to confine some of the cores that are twice as strong as the internal kinetic, thermal and magnetic energies that act to provide support.  The formation of a sheet in these simulations leads to an enhanced thermal surface pressure ($\rho c_s^2$) on the cores within it as the density in the sheet is higher, and the motions of the fluid along the field lines leads to an enhanced dynamic pressure out of the sheet.  

The simulations with intermediate values of the critical parameter $\Gamma$ have many bound cores that are significantly out of equilibrium, compared to the highly supercritical runs.  These cores are strongly confined both by surface effects and by gravity.

\subsection{Kinetic energy evolution}
We plot the kinetic energy $E_k$ (normalized to their initial values at $t=0$, $E_{k0}$) of the simulations as a function of time (in units of the flow-crossing time $t_\mathrm{flow}$, the time it takes for a shock moving at the RMS velocity to cross the simulation domain) in Fig. (\ref{fig_decay}).  Our results agree with previously reported results that the kinetic energy decays as $t^{-1}$, marked by the solid straight line in Fig. (\ref{fig_decay}) \citep{maclow_etal98,stone_ostriker_gammie98,biskamp_mueller99,padoan_nordlund99,ostriker_stone_gammie01,cho_lazarian03}. In all cases, collapse occurs after $\sim 1.5-2 \;t_\mathrm{flow}$, accompanied by an increase in the kinetic energy due to the presence of supersonic infall to the most massive cores.  At this point the gravitational energy of the simulation dominates over the kinetic, thermal and magnetic energies.

The energy spectrum for Run E14b is also plotted in Fig. (\ref{fig_decay}) at the time of the analysis.  This run began with a $k^{-2}$ power spectrum, but the kinetic energy evolved to a Kolmogorov spectrum over the inertial range, although there is an excess of energy on large scales.  In comparison with the initial state, we see that substantial damping from numerical dissipation in the smallest scale (largest $k$) modes has taken place by the end of the simulation. 

\subsection{Effect of fields on star formation efficiency}
In TP04, we estimated an upper limit to the star formation efficiency (SFE) of the simulations, where the SFE is defined as the ratio of the mass contained within bound cores to the total mass in the simulation.  The SFE was 40-50 per cent for simulations with 4.6 Jeans masses, and higher for simulations with fewer Jeans masses.  This is comparable to the estimates that \citet{clark_bonnell04} found for unmagnetized, unbound clouds.

The SFE for each of the simulations discussed in this paper are presented in Table \ref{table_sfe}.  \citet{vs_kim_bp05} noted that the SFE systematically decreased as the magnetic field strength increased.  We see some evidence for higher SFE with larger $\Gamma$, supporting this trend, but with notable outliers for Run E14b (with notably small SFE, possibly related to the greater levels of initial turbulence in this simulation) and Run C5e.
\citet{passot_etal95} report a similar trend for higher SFE at intermediate field strengths, which they attribute to magnetic braking acting to reduce the local shear that could otherwise disrupt a core.  The magnetic field channels the flow into the sheet; when the sheet fragments gravitationally, a much larger fraction of the mass is accumulated in self-gravitating cores.  Only at very large field strengths does the field retard the collapse in these sheets.

Our definition for the SFE implicitly assumes that a significant fraction of the gas in a core will collapse as a part of the protostar.  However, there are several physical processes that can occur in these later stages that our model does not take into account, some or all of which will occur in a cluster-forming region like that represented by our simulations.  Ultraviolet radiation and ionization from the most massive protostars will heat the gas, increasing the Jeans mass and thermal support and reducing the accretion rate \citep{franco_etal94}.  Stellar winds can blow accreting material away from the protostar \citep{adams_fatuzzo96}.  Magnetically-driven outflows can return gas to the cloud \citep{pudritz_norman86,shu_etal88,krumholz_etal05}.  All of these processes will serve to decrease the SFE we estimate.  However, most of the material that was originally in the core probably ends up in a star via accretion through a disc.

There are two further limitations to our calculation of the star formation efficiency that arise from the numerical limitations of our model.  First of all, many of the cores will continue to accrete gas after the time the simulations stopped due to violation of the Truelove criterion.  This will lead to an increase in the estimated values of the SFE.  Furthermore, our periodic boundary conditions artificially prevent scattering of cores out of the cloud and enhance the likelihood of interactions (e.g. \cite{li_etal04}).  Allowing the cores to disperse away from one another into lower-density gas would serve to further decrease the SFE (e.g. the work of \cite{bate_bonnell_bromm02}).

\section{Distribution Functions of Fluctuations and Cores}\label{section_distribution}

In this section, we provide the complete set of statistical 
properties of cores that form in our simulations.  
These include the distribution of core masses, angular momenta, 
spins ($\Omega$), magnetizations ($\beta$), radii, and mean densities.  
These distribution functions are calculated at the end of our simulations, 
which are determined by the times at which we can no longer 
adequately follow the collapse of the densest core due to the 
violation of the Truelove criterion (ie, the local Jeans length
can no longer be adequately resolved).
The precise time at which this occurs varies from simulation to simulation.

\begin{figure*}
\includegraphics[width=168mm]{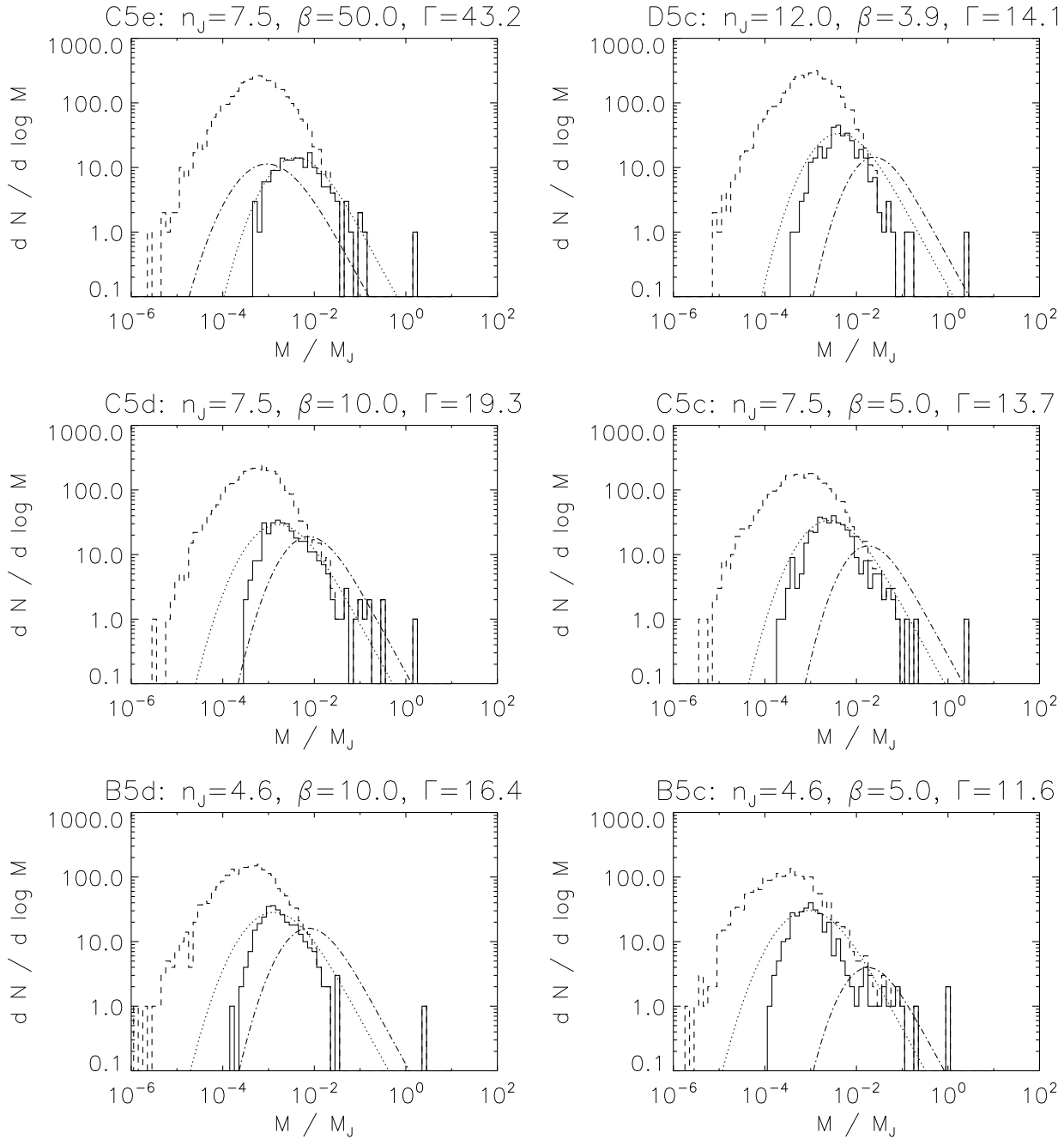}
\caption{Distribution of core masses at the end of each of the simulations.  The distribution marked by a solid line contains only those cores that are gravitationally bound; the distribution of all cores is marked by a dashed line.  The dot-dashed line is the predicted spectrum of Padoan \& Nordlund, using our initial conditions; the dotted line is a fit to the same spectrum to the bound core distribution.\label{fig_mass}}
\end{figure*}

\begin{figure*}
\includegraphics[width=168mm]{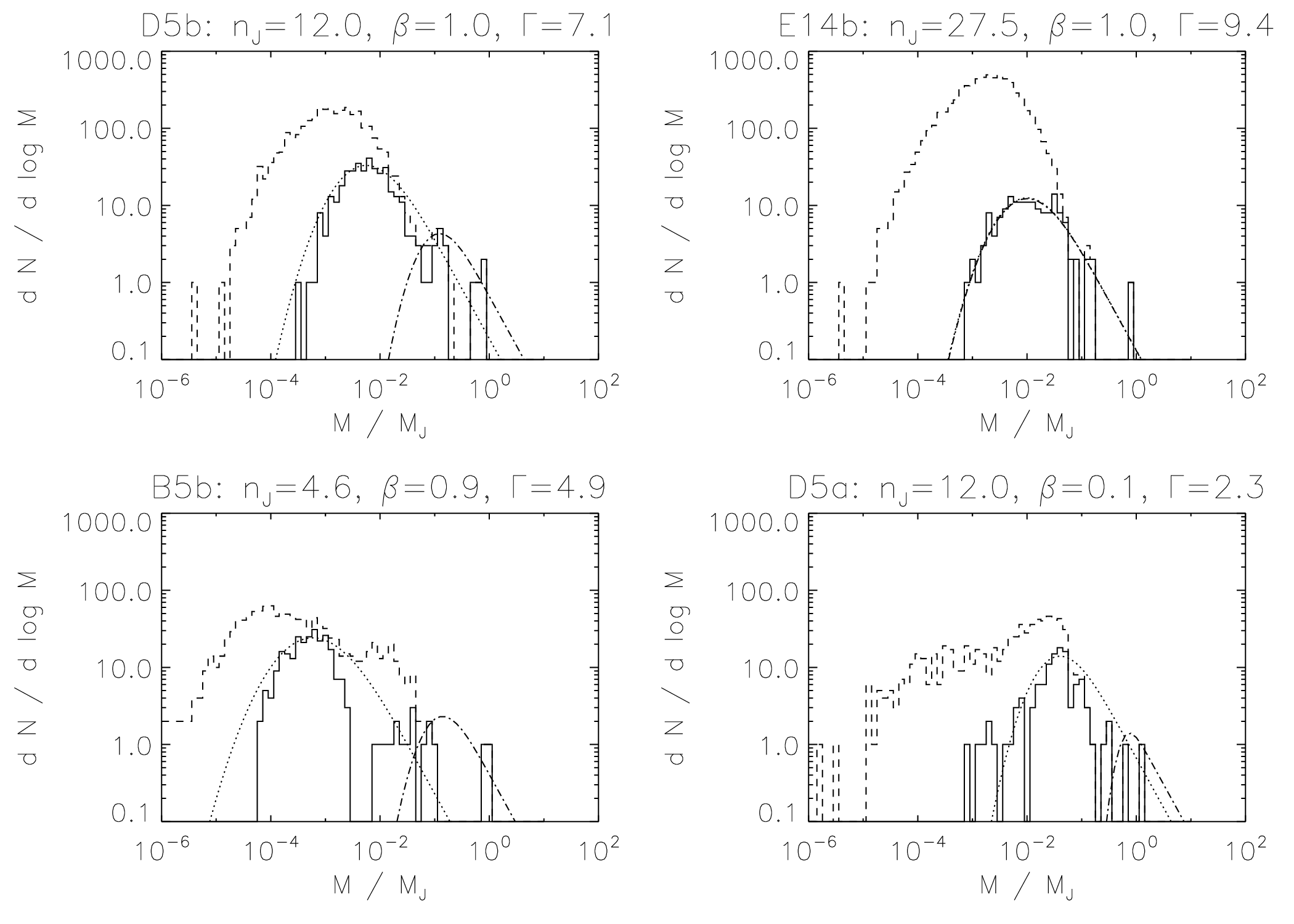}
\contcaption{}
\end{figure*}

\begin{table}
\begin{tabular}{lrrrr}
\hline Run & $n_J$ & $\Gamma$ & $\frac{M_{peak}}{M_J}$ (measured) & $\frac{M_{peak}}{M_J}$ (predicted) \\
\hline 
B5b& 4.6 & 4.9 & 6.3 x $10^{-4}$ & 0.13\\
B5c& 4.6 & 11.6 & 7.9 x $10^{-4}$ & 0.020\\
B5d& 4.6 & 16.4 & 1.3 x $10^{-3}$ & 7.9 x $10^{-3}$\\
C5c& 7.5 & 13.7 & 2.5 x $10^{-3}$ & 0.020\\
C5d& 7.5 & 19.3 & 1.6 x $10^{-3}$ & 7.9 x $10^{-3}$\\
C5e& 7.5 & 43.2 & 4.0 x $10^{-3}$ & 1.0 x $10^{-3}$\\
D5a& 12.0 & 2.3 & 0.040 & 0.79\\
D5b& 12.0 & 7.1 & 5.0 x $10^{-3}$ & 0.13\\
D5c& 12.0 & 14.1 & 4.0 x $10^{-3}$ & 0.025\\
E14b& 27.5 & 9.4 & 0.010 & 0.010\\
\hline
\end{tabular}
\caption{The mass corresponding to the peak of the mass distributions measured in Fig. \ref{fig_mass}, and the corresponding mass predicted from the Padoan-Nordlund model using our initial conditions.\label{table_peakmass}}
\end{table}

\begin{figure*}
\includegraphics[width=168mm]{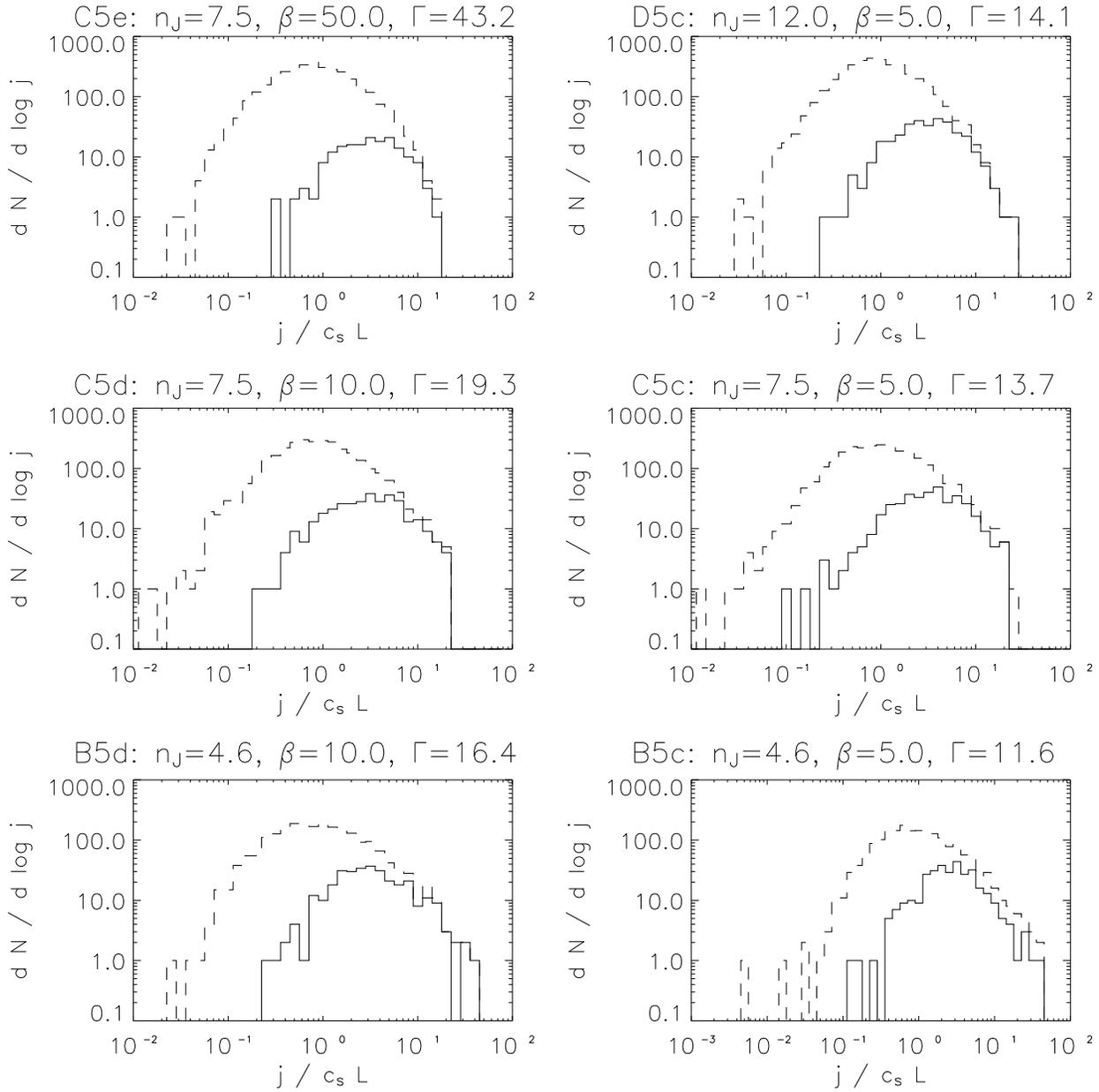}
\caption{Distribution of core specific angular momenta.\label{fig_j}}
\end{figure*}

\begin{figure*}
\includegraphics[width=168mm]{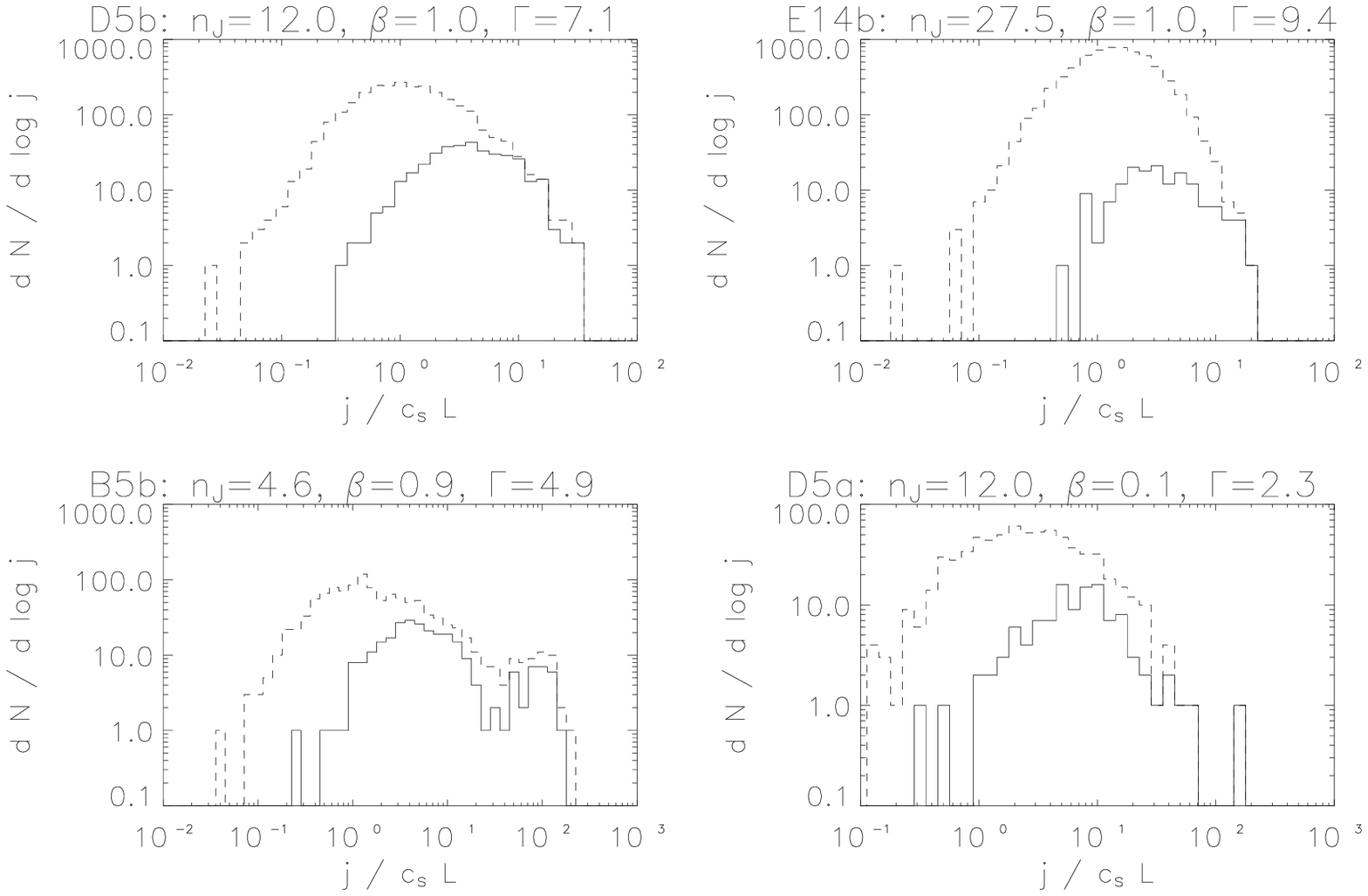}
\contcaption{}
\end{figure*}

\begin{figure*}
\includegraphics[width=168mm]{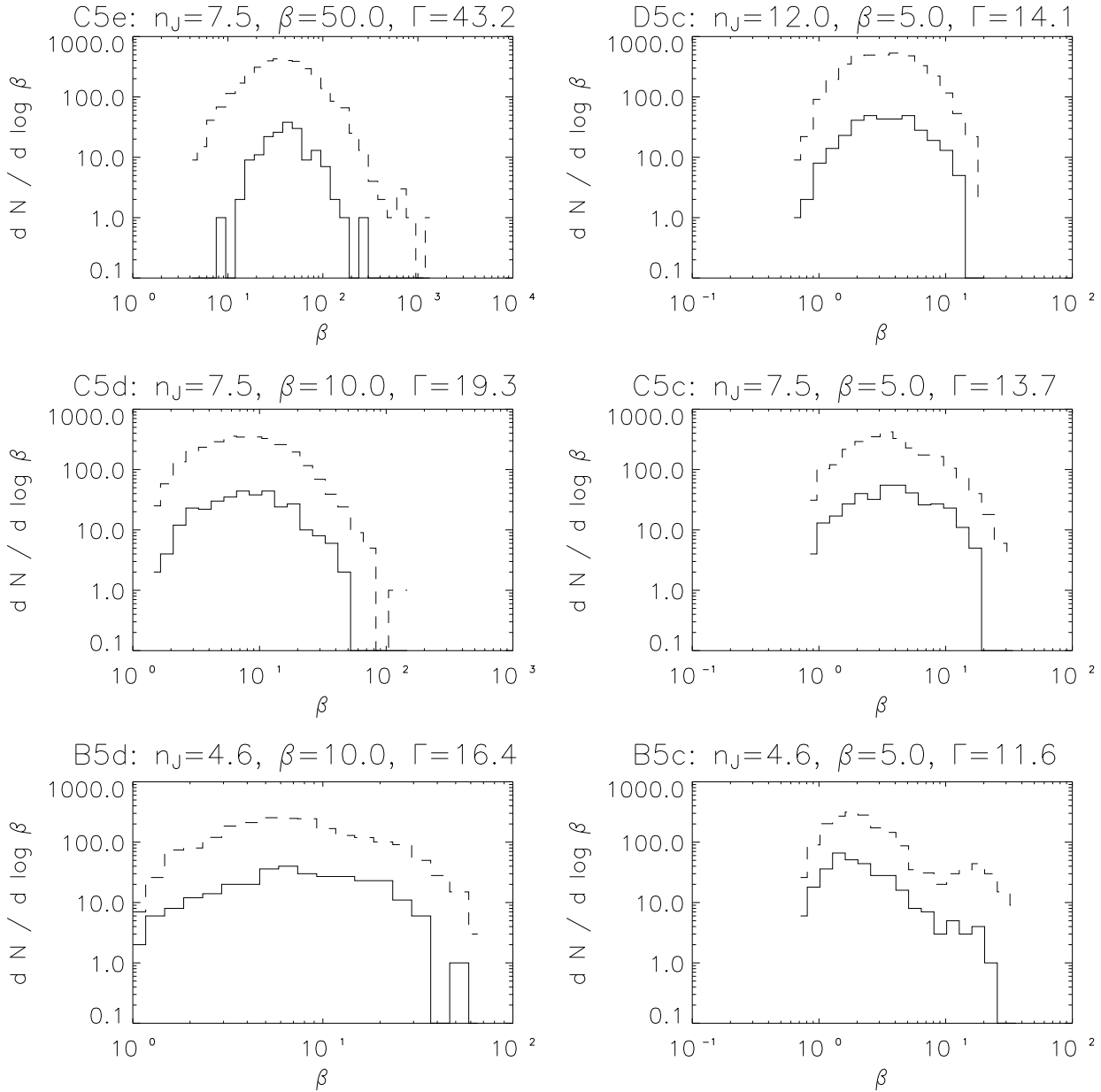}
\caption{Distribution of mean core $\beta$.\label{fig_beta}}
\end{figure*}

\begin{figure*}
\includegraphics[width=168mm]{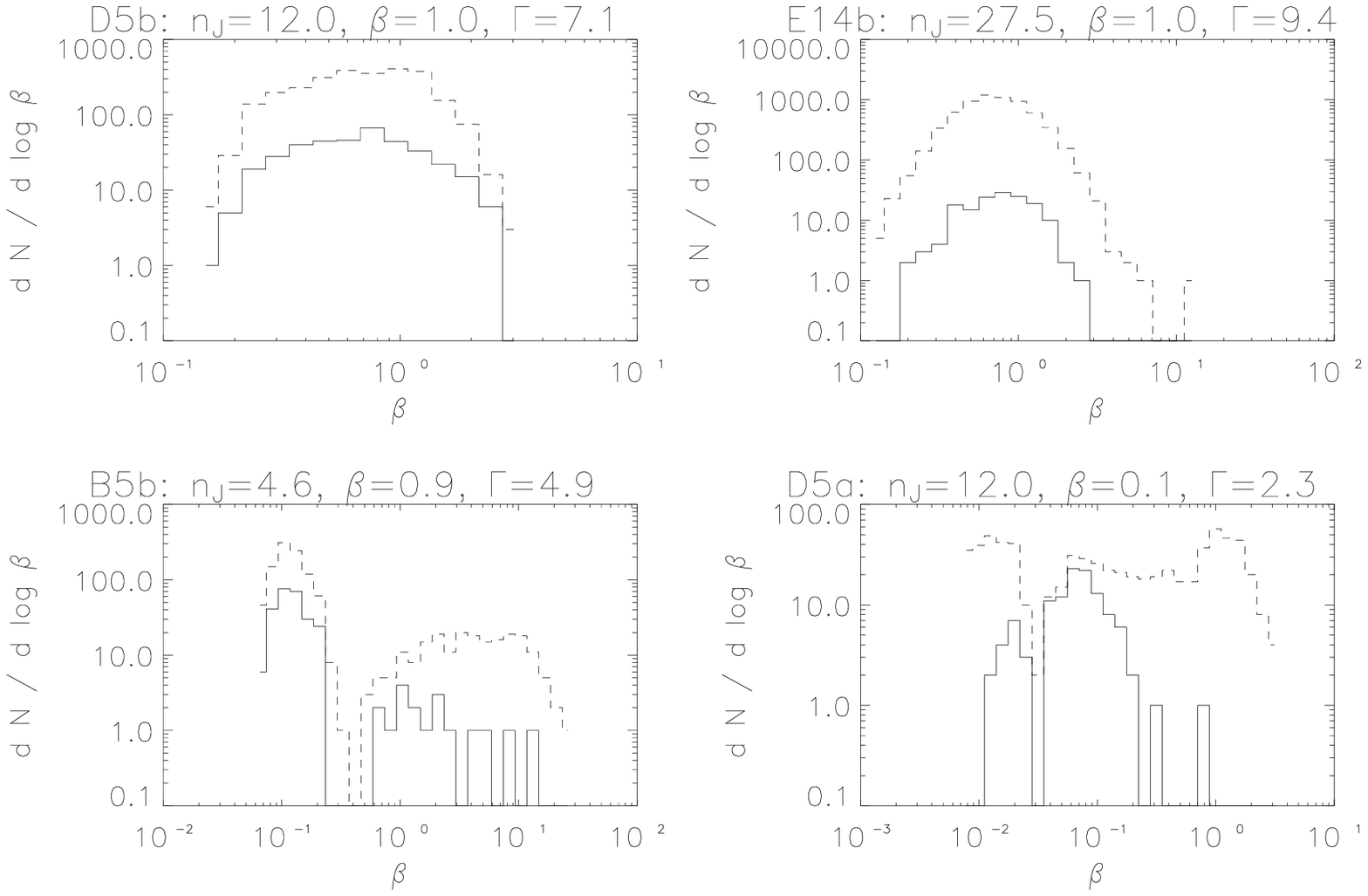}
\contcaption{}
\end{figure*}

\begin{figure*}
\includegraphics[width=168mm]{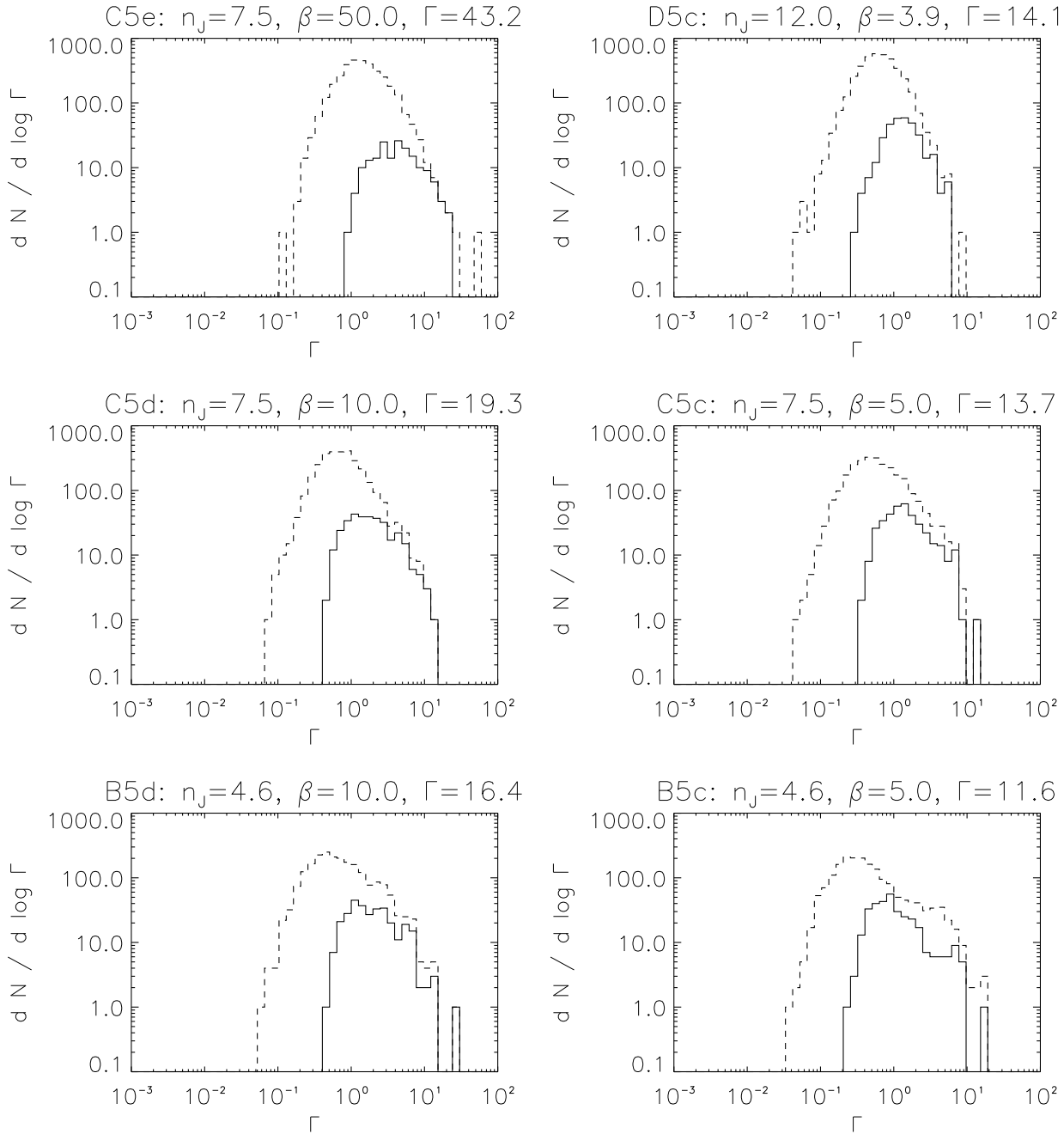}
\caption{Distribution of the mass-to-flux ratio $\Gamma$ of cores.\label{fig_gamma}}
\end{figure*}

\begin{figure*}
\includegraphics[width=168mm]{fig13b.eps}
\contcaption{}
\end{figure*}

\begin{figure*}
\includegraphics[width=168mm]{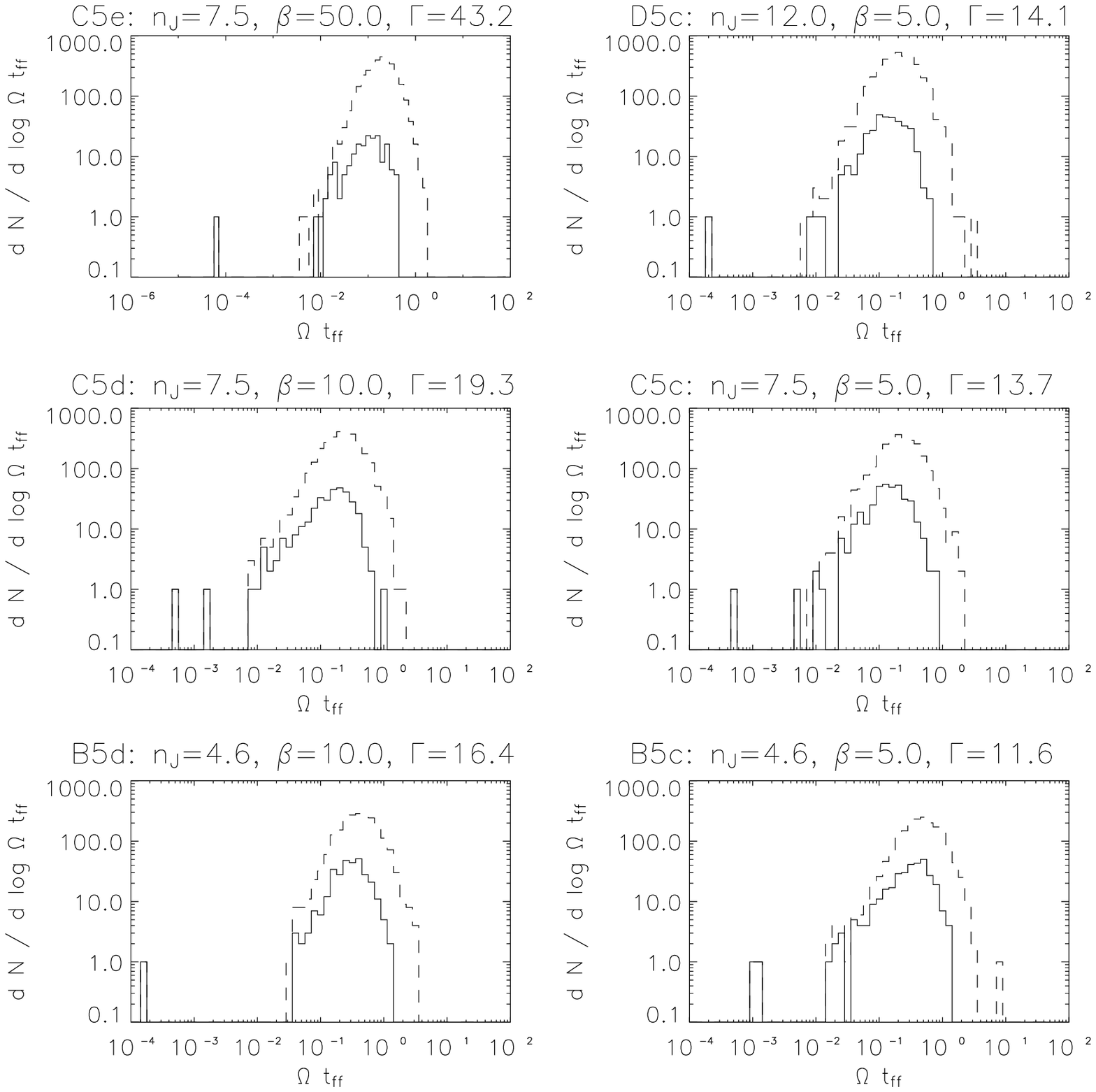}
\caption{Distribution of $\Omega t_\mathrm{ff}$.\label{fig_omtff}}
\end{figure*}

\begin{figure*}
\includegraphics[width=168mm]{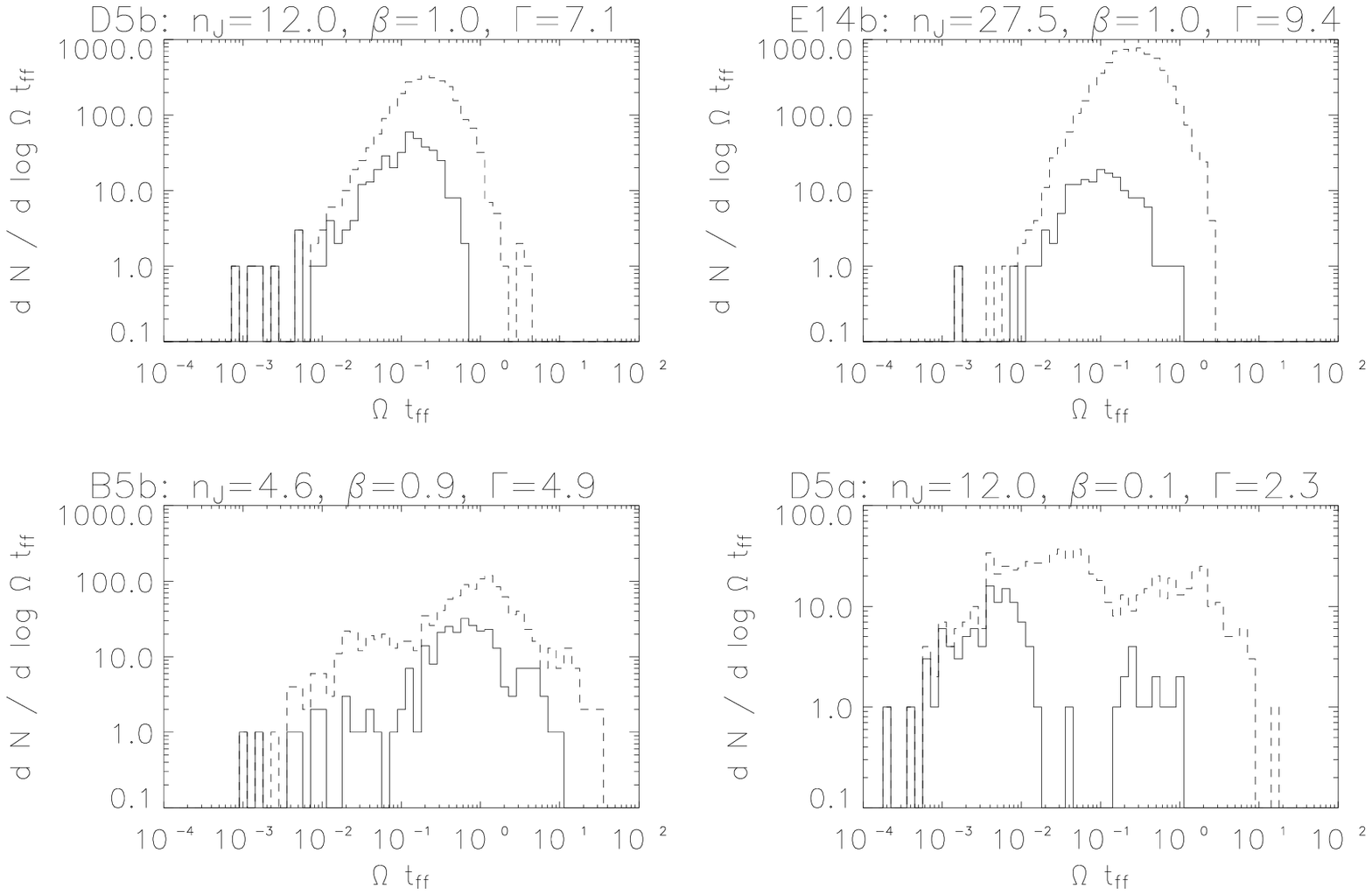}
\contcaption{}
\end{figure*}

\begin{figure*}
\includegraphics[width=168mm]{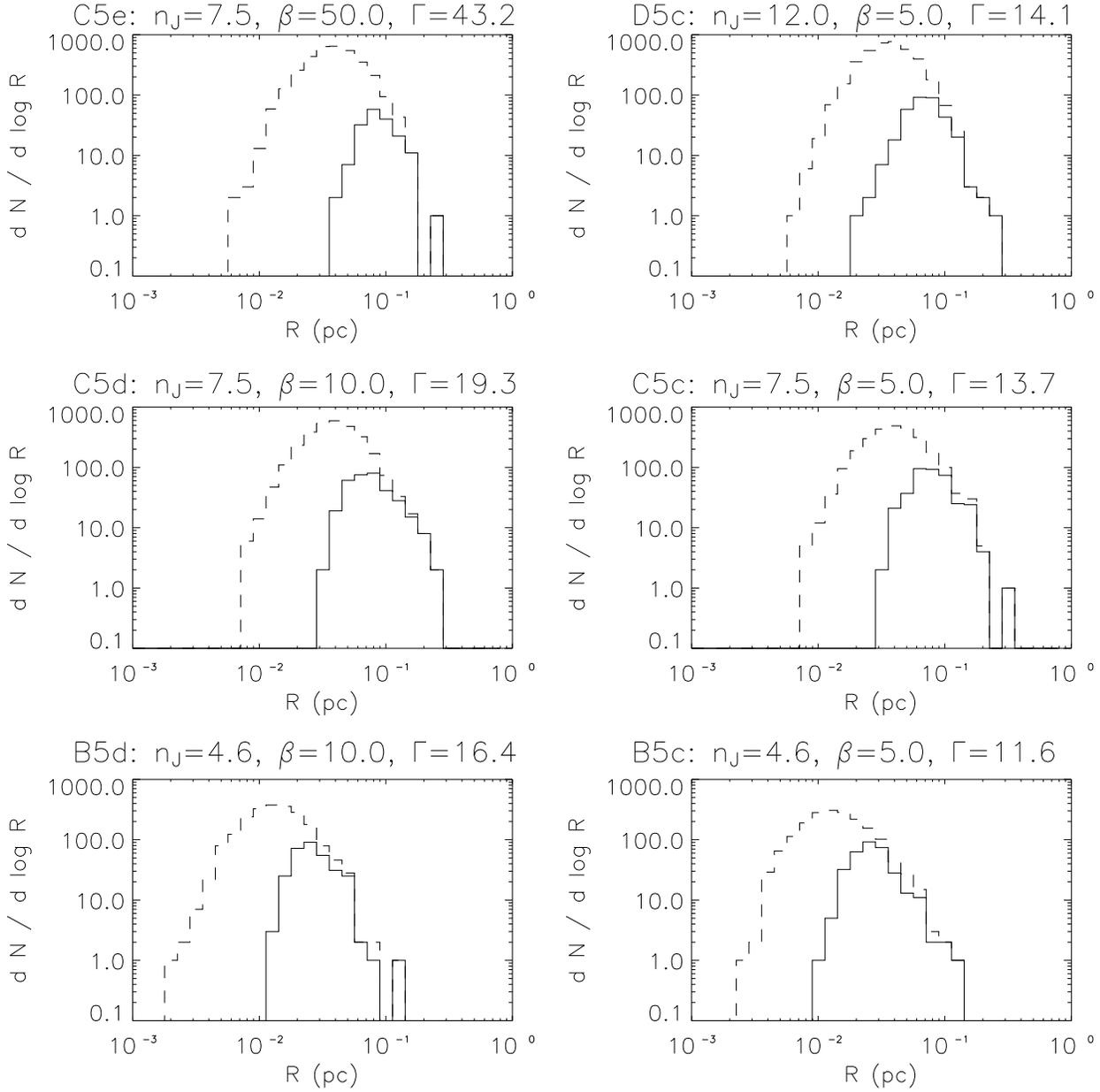}
\caption{Distribution of core radii\label{fig_radii}.  These plots are scaled to the initial values in Table \ref{table_ic}.  1 zone corresponds to $10^{-3}\;\mathrm{pc}$ for Runs B5b, B5c and B5d; 1 zone corresponds to $4\times10^{-3}\;\mathrm{pc}$ for the other runs.}
\end{figure*}

\begin{figure*}
\includegraphics[width=168mm]{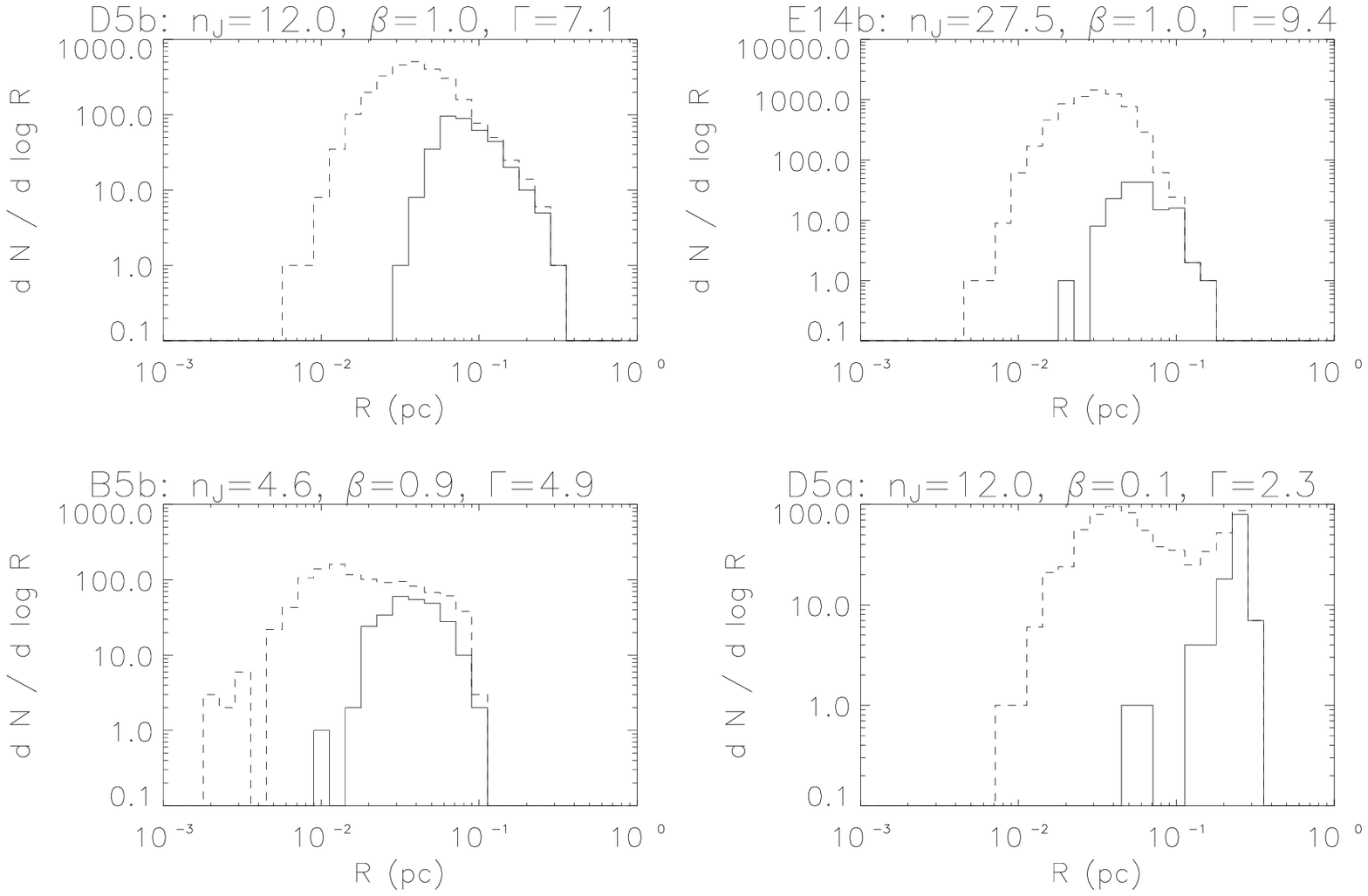}
\contcaption{}
\end{figure*}

\begin{figure*}
\includegraphics[width=168mm]{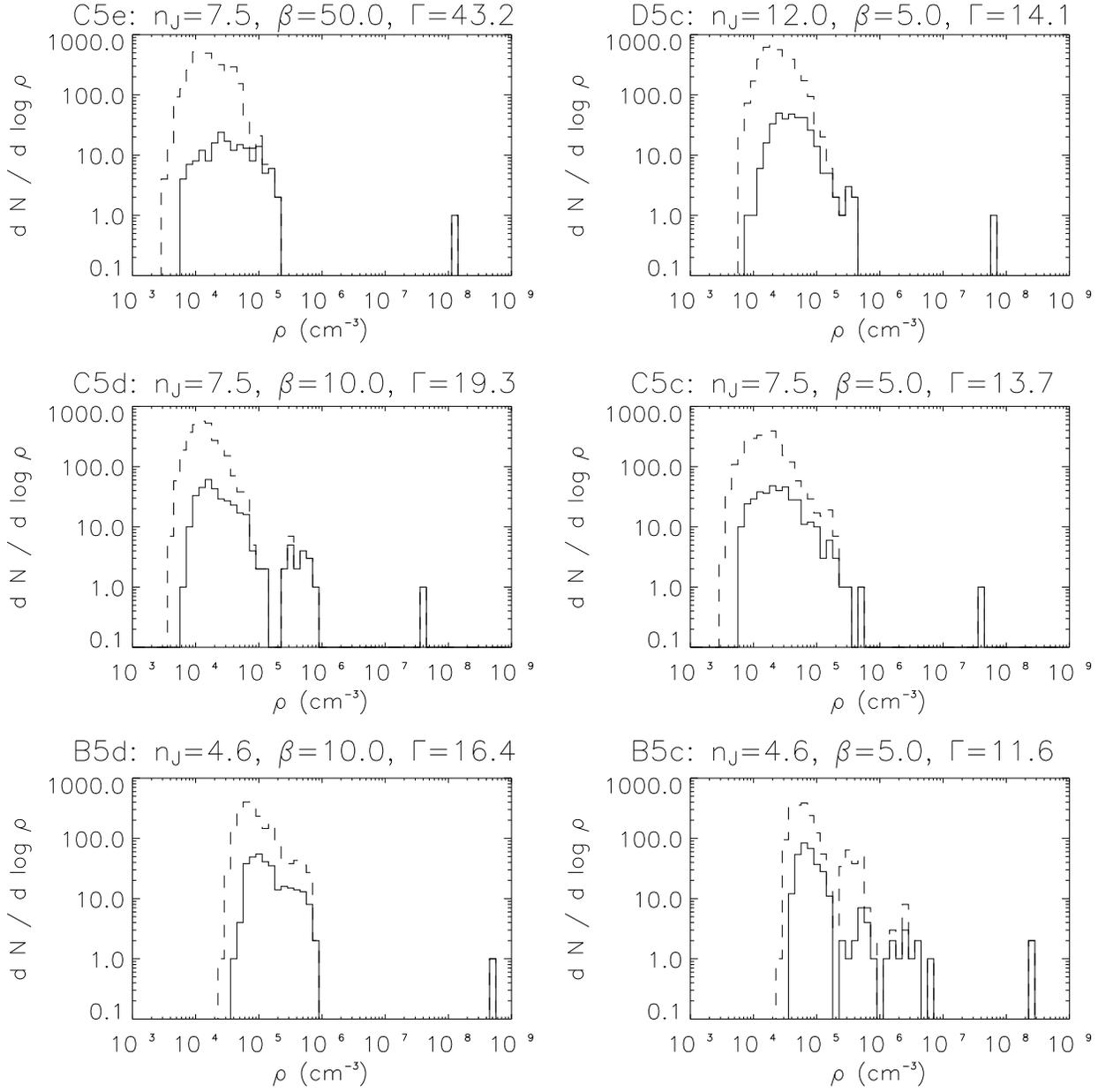}
\caption{Distribution of the peak core density, scaled to the initial values in Table \ref{table_ic}.\label{fig_dens}}
\end{figure*}

\begin{figure*}
\includegraphics[width=168mm]{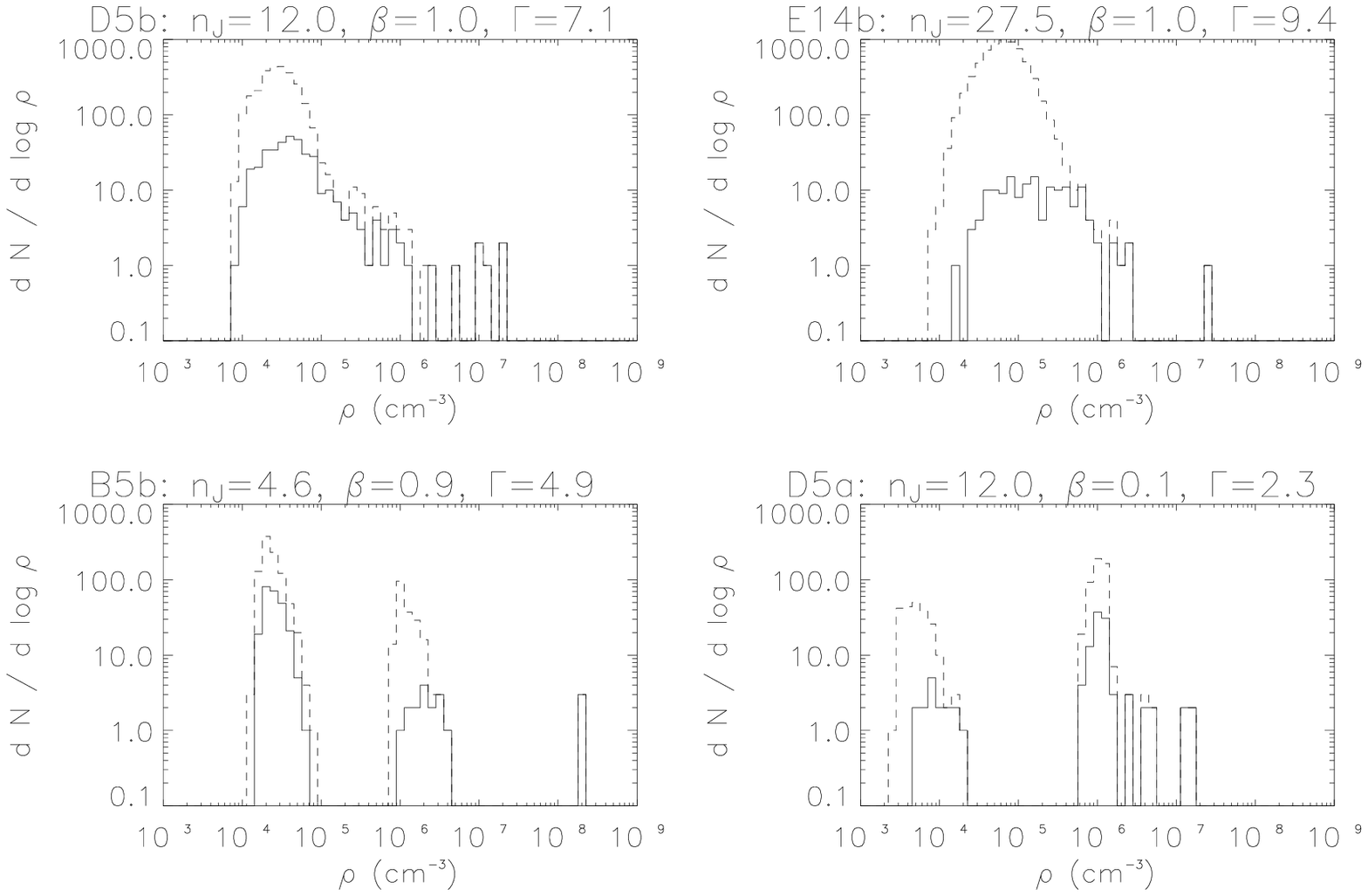}
\contcaption{}
\end{figure*}

\subsection{Mass Distribution}

We have plotted the distribution of masses for all of the condensations within our simulations in Fig. (\ref{fig_mass}).  In all cases, we have plotted the set of all condensations with a dashed line, and the gravitationally bound cores only in solid lines.  The bound core distribution generally consists of a power-law behaviour at high masses, with a turnover at $10^{-3}-10^{-2} m_J$ (when scaled to the initial values in Table \ref{table_ic}, this corresponds to $0.1-0.4 m_\odot$).  This type of behaviour is consistent with other results reported in the literature (e.g. \cite{gammie_etal03} has a turnover at $0.5 m_\odot$; \cite{li_etal04} find a turnover at $\sim 1-2 x 10^{-2} m_J$; \cite{bate_bonnell05} has the turnover at $0.07 m_J$), and consistent with observations of low-mass star forming regions \citep{luhman00,luhman_etal00,luhman_etal03}.  We produce bound cores with masses an order of magnitude below this peak, and thus below the hydrogen-burning limit, as suggested should be the case for supersonic turbulence \citep{padoan_nordlund04}.

  We attempted to fit the bound core mass distribution with the theoretical distribution of \citet{padoan_nordlund02}.  This distribution posits that the high-mass part of the spectrum is a power law function that arises from the origin of the cores in shocks.  There is a turnover at low masses due to the fact that even the shocks cannot compress small enough fluctuations to the point where they become Jeans unstable.  The Padoan-Nordlund distribution has the form
\begin{eqnarray}
\frac{d N(m)}{d \log{m}} & \propto & m^{-1.29} \; \mathrm{erf}\left[\frac{\sqrt{2}}{\sigma}\ln(m) + \sqrt{2}\sigma\right]
\end{eqnarray}
where the power-law -1.29 exponent is derived from the exponent
$\gamma$ of the turbulent power spectrum (in our case, $\gamma=-5/3$),
via the equation $-3/(4+\gamma)$ (see \cite{padoan_nordlund02}).  This
exponent arises from the density jump in a magnetized, isothermal
shock, and an assumption that the number of cores in a given volume
scales in a self-similar manner with the volume. $\sigma$ is a measure
of the width of the density probability distribution function; in the
turbulence model used by \cite{padoan_nordlund02}, it is related to
the Alfv\'en Mach number via $\sigma^2 \sim
\ln(1+\frac{1}{4}\mathcal{M}_A^2)$.  The error function arises from
the requirement that the mass of any given turbulence fluctuation be
greater than the local Jeans mass in order for collapse to proceed,
and leads to a turnover at low masses.  As the likelihood of a
fluctuation being sufficiently compressed in a shock so that it
becomes Jeans unstable is greater if the turbulence is stronger (and
thus the stronger shocks can compress the fluid more), the width of
this distribution is directly related to the Mach number of the
turbulence.  We fit (dotted line in Fig. \ref{fig_mass}) this
distribution to our bound core mass function, allowing the local Mach
number and the height to vary.  For comparison, we also provide the
predicted spectrum of the PN02 model using the initial Alfv\'en Mach
number - a curve that is indicated by a dot-dashed line.

We find that in most of our simulations, the {\it shape} of the
distribution of bound core masses is reproduced well by the
Padoan-Nordlund spectrum.  As is readily seen by comparing these
curves, their model does not fit the peak mass of the distributions
with one exception (E14b), as demonstrated in Table
\ref{table_peakmass}.  The greatest discrepancy is that the position
of the peak of the mass spectrum in our simulations is independent of
the initial Alfv\'en Mach number, contrary to the predictions of
Padoan \& Nordlund.  We also find that the distribution of cores
produces far more low-mass cores than would be predicted by the
Padoan-Nordlund model for our initial Alfv\'en Mach numbers.  The
trend towards lower masses could in part be explained by the important
role played by the surface pressure in binding the cores (see Fig.
\ref{fig_virial}), which could allow smaller fluctuations to remain
overdense long enough for gravitational collapse to begin.

In the non-magnetic simulations of TP04, as soon as one core began to collapse, it collapsed to the point that the simulation could no longer follow it due to violation of the Jeans criterion.  It did so before any other core was able to evolve to a significant degree, and thus there was always one high-mass outlier in the mass distributions.  The same effect can be seen in these simulations that are highly supercritical; the first core to collapse does so quickly enough that it alone accretes a significant portion of the mass before the simulation is halted.  In the runs with a stronger magnetic field (and are thus less supercritical), this effect is much less pronounced and can even disappear completely.  

Our initial conditions are similar to that of \citet{li_etal04}, who find a similar result for the mass spectrum of bound cores.  As these authors utilize a more restrictive definition of a core (in that they don't allow for fluctuations that are bound by virtue of their surface pressure), the correspondence between our results and theirs suggests that we are not biasing our results by using a broader definition of a core.  \citet{li_etal04} examine the evolution of the mass distribution and find that it becomes shallower with time as cores merge.  We expect to find a similar pattern here, although we do not examine it further.

\subsection{Angular Momentum Distribution}

Fig. (\ref{fig_j}) plots a histogram of the distribution of specific angular momenta of the cores in each simulation.  As in Fig. (\ref{fig_mass}), the solid lines mark the distribution for the bound cores, and the dashed lines the distribution of all fluctuations, bound and unbound.  There does not appear to be any clear relationship between the peak of the angular momentum distributions of the cores in Fig. (\ref{fig_j}) with either $n_J$ or $\beta$.  In all of the simulations, the core angular momentum distribution peaks at $\sim 5-10 c_s L$, while the unbound condensation angular momentum distribution peaks at the slightly lower values of $\sim 1-3 c_s L$.  (For comparison, $c_s L \sim 10^{22}-10^{23}\;\mathrm{cm^2 s^{-1}}$ for our initial conditions.)  This is comparable to the results of $\sim 4\times10^{22}\;\mathrm{cm^2 s^{-1}}$ of \citet{gammie_etal03}, and significantly larger than the $2\times10^{21} \;\mathrm{cm^2 s^{-1}}$ that \citet{burkert_bodenheimer00} find from simulated observations of turbulent noise in an otherwise static core.  It is also several orders of magnitude larger than the angular momenta of prestellar cores found in \citet{jappsen_klessen04}, although in that case the difference is likely due to the fact that they define cores as collapsed objects represented by sink cells, and thus exist on a much smaller scale than what we can resolve.  Observed velocity gradients in molecular cloud cores have been interpreted as rotation, with estimate specific angular momenta of $j\sim 10^{21}-10^{22}\;\mathrm{cm}^{-3}$ \citep{goodman_etal93,caselli_etal02b}, slightly less than our estimates.  We note that the specific angular momenta of our cores decreases with radius, which we will describe in more detail in a future paper.  

We see a bimodal angular momentum distribution for run B5b, similar to the bimodal mass distribution.  There is not a clear explanation for this, although it too is likely related to the differences in the mechanics of the collapse in a sheet versus collapse in the envelope.

\subsection{Magnetization ($\beta$) and Mass-to-Flux ($\Gamma$) Distributions}

\begin{table}
\begin{tabular}{lll}
\hline Run & $\beta$ of Simulation & Median $\beta$ of cores \\
\hline
B5b & 0.9 & 0.34 \\
B5c & 5.0 & 3.2 \\
B5d & 10.0 & 10.9 \\
C5c & 5.0 & 5.3 \\
C5d & 10.0 & 11.4 \\
C5e & 50.0 & 53.9 \\
D5a & 0.1 & 0.090 \\
D5b & 1.0 & 0.87 \\
D5c & 3.9 & 4.5 \\
E14b & 1.0 & 0.91 \\
\hline
\end{tabular}
\caption{Initial $\beta$ of each simulation and the median $\beta$ of the cores produced in each simulation.\label{table_beta}}
\end{table}

The mean magnetization ( $\beta$) of each core for our standard 
set of simulations is plotted in Fig. (\ref{fig_beta}).  As in Fig. (\ref{fig_mass}), the solid line marks the distribution for the bound cores that were identified in that simulation, and the dashed line marks the distribution of all fluctuations.  The median value of $\beta$ for the bound cores is within $15\%$ of the original value of $\beta$ in all of the simulations except B5b and B5c (see Table \ref{table_beta}); in B5b we see a large number of cores with very small values of $\beta$ that is separate from the distribution centred on the mean $\beta$ of 0.9, and similarly in B5c we see that the distribution of $\beta$ is strongly skewed towards small values, unlike what we see in the other simulations.

The distribution of the mass-to-flux ratio of the cores is plotted in Fig. (\ref{fig_gamma}), with solid lines denoting bound cores and dashed lines denoting all fluctuations.  We see that nearly all of the bound cores that are produced have mass-to-flux ratios less than the initial mass-to-flux ratio of the original simulation domain.  We also find that significant numbers of subcritical cores can be produced, even in the simulations that start in a highly supercritical state.  This important result suggests that the observed near-critical mass-to-flux ratios of \citet{crutcher99} can develop from a cloud that is initially supercritical. Thus, it is possible to avoid the sheet-like collapse behaviour of an initially near-critical cloud while still producing near-critical cores similar to those seen in star-forming regions.

In the low-$\Gamma$ runs, the separation between the cores in the high-density sheet and low-density atmosphere is especially pronounced in Figs. (\ref{fig_beta}-\ref{fig_gamma}).  As the collapse into these sheets preferentially occurs along field lines, the magnetic field strength both inside and outside of the sheet is approximately constant; since $\beta \propto \rho/B^2$, the density contrast between sheet and environment is directly reflected in the $\beta$ distribution for these marginally supercritical runs.  It is apparent that in these marginally supercritical simulations, the magnetic field is not significantly compressed.

\subsection{$\Omega t_\mathrm{ff}$ Distribution -- Upper Limits on the Binary Frequency?}

A quantity that has proved useful in determining whether a collapsing core fragments is the product of $\Omega$, the mean angular rotation rate of the core, and $t_\mathrm{ff}$, the free-fall time at the centre of the core \citep{matsumoto_hanawa03,banerjee_etal04}.  We calculate this product using the specific angular momentum, central density and mean radius of each condensation:
\begin{eqnarray}
\Omega t_\mathrm{ff} & = & \frac{j}{R^2}\sqrt{\frac{3\pi}{32 G \rho}}
\end{eqnarray}
The results are plotted in Fig. (\ref{fig_omtff}).

For most of the simulations, $\Omega t_\mathrm{ff}$ is around 0.1.
The collapse of such objects will produce discs, which
\citet{matsumoto_hanawa03} and \citet{banerjee_etal04} suggest will
fragment into a combination of rings and bars, depending on the amount
of anisotropy.  Cores with $\Omega t_\mathrm{ff} \le 0.03$ tend to
collapse smoothly into discs without significant fragmentation, but
very few of our cores fall into this category.  Cores with $\Omega
t_\mathrm{ff} > 0.3$ will not collapse beyond some minimum radius,
where rotation will cause the core to expand again.  About $ 20\%$ of
the cores in our simulations fall into this last category.  We see
some notable exceptions for simulations with low $\Gamma$.  For Run
B5b, $\Omega t_\mathrm{ff}$ tends to be higher as the sheet that forms
produces a number of cores with very low central densities and masses
away from the sheet; the net result is a significantly larger
free-fall time.  Conversely, run D5a has extremely low values of
$\Omega t_\mathrm{ff}$ as the bound cores generally have extremely
large radii (Fig. \ref{fig_radii}), thus resulting in a low mean
rotation rate.  These distributions of $\Omega t_\mathrm{ff}$
therefore say something very interesting about the binary fraction
expected in young star clusters.  Our data suggest that it should be
rather high, in agreement with \citet{bate_bonnell05} and
\citet{dd_etal04b}.

\subsection{Radius and Central Density Distributions}
In Fig. \ref{fig_radii} we plot the distribution of the radii of the cores found in our simulations, scaled to the default values given in Table \ref{table_ic}.  In general, the bound core radius distribution peaks at 10\% of the box size of the simulation.  As a result, most of the cores produced by these simulations are 0.02-0.2 pc in size.  This is the same scale as where most of the cores in the \citet{jijina_etal99} dataset are found.  Only a very tiny fraction of cores, $\sim 1$\%, are resolved with fewer than 10 zones.  In contrast, 10-20\% are typically resolved with at least 30 zones.

The distribution of the peak density of each core is presented in Fig. \ref{fig_dens}.  The existence of a core that is undergoing runaway accretion is evident in all of the highly supercritical simulations, as this core has a significantly higher density than any of the rest.  This is less of an effect in the stronger-field simulations, due to the reduced compressibility of the gas from the presence of the strong magnetic fields \citep{passot_etal95,balsara_etal01a,heitsch_etal01,vs_etal05}.  The peak of these distributions are between $10^4-10^5\;\mathrm{cm}^{-3}$, densities that are well-traced by ammonia maps (although it should be noted, we chose our initial conditions so that this would be true) and again agree with \citet{jijina_etal99} who characterize bound cores through ammonia surveys.


\section{Discussion and Conclusions}

We have presented a detailed examination of the physical properties of cluster forming
cores as they arise in simulations of decaying turbulence in magnetized, turbulent 
"clumps" within molecular clouds.  This effort is aimed at trying to understand the
origin of the core mass function, as well as many other properties of cores and eventually,
the stars that form within them.    
In pursuing this goal, we have been careful to examine all of the physical
forces that are acting upon the fluctuations in our simulations - 
our virial analysis of cores is the most exhaustive analysis we can find in the literature.
Our results are also not sensitive to the nature of the initial spectrum of the velocity field
fluctuations that is imposed on the simulations.

We find that an initial turbulent, 
magnetized molecular cloud must be significantly supercritical 
if its subsequent fragmentation is to resemble at all the properties of evolved molecular cloud cores 
that we see today.  The cores that are produced in these highly supercritical 
simulations can still have (local) mass-to-flux ratios that are close to critical, 
reproducing the trends seen by \citet{crutcher99} for observed cores.  
Simulations that are not initially significantly supercritical preferentially collapse 
along field lines to form sheets, and the fragmentation of these sheets produce much flatter mass distributions.

We find through the use of the virial theorem that the 
surface pressure and surface magnetic field play a critical role 
in creating bound cores (see also \cite{bp_vs_scalo99}). 

We obtain a distribution of masses for the bound cores that 
shows a power-law at large masses for simulations that began with 
moderately supercritical magnetic fields.  The peak of these mass 
distributions are robustly found at 0.001-0.01 times the initial Jeans mass of the simulation.  
We fit these core mass distributions with the theoretical mass spectrum of \citet{padoan_nordlund02}  
and generally obtain reasonable fits to the {\it shape} of the core mass
spectra using their model (albeit with fewer very-low-mass cores produced 
than expected by the Padoan-Nordlund mass spectrum).  Cores that are only marginally supercritical 
have a significantly flatter mass distribution, and do not appear to develop a clear power-law at large masses.  
However, the {\it peak mass} of the core mass spectra in our simulations are not well fit by this model. 
In particular, there is little sensitivity of the peak mass in our simulations to 
the initial Mach number of the turbulence
as proposed by their model. 

The bound cores that are produced have specific angular momenta that, 
when scaled to appropriate physical units, are on the order 
of $10^{22}-10^{23}\;\mathrm{cm}^2\mathrm{s}^{-1}$.  On the basis of these 
spin rates and the detailed collapse calculations of 
\citet{matsumoto_hanawa03} and \citet{banerjee_etal04}, many of these cores 
are likely to further fragment during the course of their collapse, likely leading to the formation of binary or multiple systems.

We find that the bound cores have a distribution of magnetization ($\beta$) that has a median 
value that is approximately the same as the mean $\beta$ of the simulation.  The mass-to-flux ratio 
($\Gamma$) of these cores, however, is generally less than the mass-to-flux ratio 
of the original simulation, by up to an order of magnitude.  This suggests that a distribution of cores can be produced that have mass-to-flux ratios that are close to critical (as seen by \cite{crutcher99}) from a cloud that is initially highly supercritical.

The distribution of core radii is a narrowly peaked function centered 
at 0.02-0.2 pc.  The distribution of core densities peaks at the 
mean density of the simulation, but is skewed towards higher densities due to 
gravitational collapse.  The marginally supercritical runs develop a double peak 
in their density distributions, with the one peak arising from the cores 
collapsing in a dense sheet and the other from cores in the low density surroundings.   

We conclude that simulations of magnetized, decaying turbulence can provide an excellent account
of the core mass functions and a wide range of core properties in such clouds.  The role of magnetic
fields can indeed be very significant in the dynamics of molecular clouds and their cluster
forming cores.  Indeed our best fits to a wide variety of observations of cloud cores suggest that the strong magnetization
of cores that is often observed is the result of compression by the turbulence, and is not
characteristic of the entire volume of the cloud.  This further suggests that turbulence, and not wide
spread cloud magnetic field, lies at the heart of the origin of core formation and the origin of the 
IMF.

We thank Chris McKee, Phil Myers and Dean McLaughlin for useful discussions.  
We also thank our referee, Dr. Mordecai-Mark Mac Low, for suggesting that we look at the mass-to-flux ratios of the 
cores, and for several useful comments throughout.  
We would like to acknowledge the SHARCNET HPC Consortium for 
the use of its computing facilities at McMaster University.  
David Tilley is funded by an Ontario Graduate Scholarship and an 
Ontario Graduate Scholarship in Science and Technology.  Ralph Pudritz is supported by the Natural Science and Engineering Research Council of Canada.

\bibliographystyle{mn2e}

\end{document}